\documentclass[draft,jgrga]{agutex}

 \usepackage[]{graphicx}
  \setkeys{Gin}{draft=false}
\authorrunninghead{MARKIDIS ET AL.}

\titlerunninghead{PIC SIMULATIONS OF MAGNETIC RECONNECTION WITH $\mathrm{O^+}$}

\authoraddr{S. Markidis,
Centrum voor Plasma-Astrofysica, Departement Wiskunde, Katholieke Universiteit Leuven, Celestijnenlaan 200B - bus 2400, B-3001 Heverlee, Belgie.
(s.markidis@gmail.com)}

\authoraddr{G. Lapenta, Centrum voor Plasma-Astrofysica,
Centrum voor Plasma-Astrofysica, Departement Wiskunde, Katholieke Universiteit Leuven, Celestijnenlaan 200B - bus 2400, B-3001 Heverlee, Belgie.}

\authoraddr{L. Bettarini, Centrum voor Plasma-Astrofysica,
Centrum voor Plasma-Astrofysica, Departement Wiskunde, Katholieke Universiteit Leuven, Celestijnenlaan 200B - bus 2400, B-3001 Heverlee, Belgie.}

\authoraddr{M. Goldman, Department of Physics, University of Colorado, Boulder, CO 80309, USA.}

\authoraddr{D. Newman, Department of Physics, University of Colorado, Boulder, CO 80309, USA.}

\authoraddr{L. Andersson, Laboratory for Atmospheric and Space Physics, Boulder, CO 80309, USA.}

\begin{document}

%
%

\title{Kinetic simulations of magnetic reconnection in presence of a background $\mathbf{O^+}$ population}
%

%
%








%
%


\begin{abstract}
Particle-in-Cell simulations of magnetic reconnection with an $\mathrm{H^+}$ current sheet and a mixed background plasma of $\mathrm{H^+}$ and $\mathrm{O^+}$ ions are completed using physical mass ratios. Four main results are shown. First, the $\mathrm{O^+}$ presence slightly decreases the reconnection rate and the magnetic reconnection evolution depends mainly on the lighter $\mathrm{H^+}$ ion species in the presented simulations. Second, the Hall magnetic field is characterized by a two-scale structure in presence of $\mathrm{O^+}$ ions: it reaches sharp peak values in a small area in proximity of the neutral line, and then decreases slowly over a large region. Third, the two background species initially separate in the outflow region because $\mathrm{H^+}$ and $\mathrm{O^+}$ ions are accelerated by different mechanisms occurring on different time scales and with different strengths. Fourth, the effect of a guide field on the $\mathrm{O^+}$ dynamics is studied: the $\mathrm{O^+}$ presence does not change the reconnected flux and all the characteristic features of guide field magnetic reconnection are still present. Moreover, the guide field introduces an $\mathrm{O^+}$ circulation pattern between separatrices that enhances high $\mathrm{O^+}$ density areas and depletes low $\mathrm{O^+}$ density regions in proximity of the reconnection fronts. The importance and the validity of these results are finally discussed.

\end{abstract}

%
%

%

\begin{article}

%
%

\section{Introduction}
The presence of heavy $\mathrm{O^+}$ ions with ionospheric origin has long been observed in Earth's magnetotail \citep{Frank:1977,Wilken:1995,Zong:1998,Mouikis:2010}. $\mathrm{O^+}$ population increases during storm times when $\mathrm{O^+}$ can even dominate $\mathrm{H^+}$ in number density and pressure \citep{Kistler:2005} and there is evidence that $\mathrm{O^+}$ species is involved in magnetic reconnection events \citep{Kistler:2005,Wygant:2005}. Besides finding the presence of $\mathrm{O^+}$ in the Earth's magnetotail during magnetic reconnection, the analysis of satellite data showed different behaviors of $\mathrm{O^+}$ and $\mathrm{H^+}$ populations \citep{Wygant:2005,Kistler:2006}. However, pure data analysis can not yet fully explain the dynamics of $\mathrm{H^+}$ and $\mathrm{O^+}$ because of the lack of resolution in space of satellite data. In support of observational analysis, theoretical and simulation studies can help first in verifying observations and second in unveiling the microphysics of magnetic reconnection in presence of mixed ion populations. In fact, simulations can be used as a microscope to study dynamics at micro-scale and to reveal the microphysics of magnetic reconnection. 

Many simulation approaches can be applied to study the effect $\mathrm{O^+}$ on the magnetosphere dynamics and on magnetic reconnection. Global simulations of the Earth's magnetosphere with $\mathrm{O^+}$ ionospheric outflow have been completed \citep{Winglee:2002,Brambles:2010,Glocer:2009,Glocer:2009b,Wiltberger:2010,Brambles:2010,Brambles:2010b}. The multicomponent fluid approach focused more specifically on the study of magnetic reconnection with $\mathrm{O^+}$ ions \citep{Shay:2004}. Moreover, the simulation of test particles moving in a preassigned electromagnetic field, that is calculated previously with a MHD simulation, allowed to study $\mathrm{O^+}$ ion trajectories in the global Earth's magnetosphere \citep{Birn:2004}. Hybrid simulations, where $\mathrm{O^+}$ and $\mathrm{H^+}$ ions are modeled with kinetic particles while electrons are treated as a fluid in a self-consistent field, have used also to study magnetic reconnection with $\mathrm{O^+}$ ions \citep{Fujimoto:1994}. Finally, a fully kinetic approach, such as the Particle-in-Cell method, where electrons are represented by particles also, has been followed by \citet{Hesse:2004, Karimabadi:2010}. 

The present paper reports the first full kinetic simulations with the physical mass ratios among electrons, hydrogen and oxygen. Previous Particle-in-Cell simulations either did not use physical mass ratios for $\mathrm{O^+}$, $\mathrm{H^+}$ and electrons \citep{Hesse:2004, Karimabadi:2010} or did not consider $\mathrm{O^+}$ ions in the magnetic reconnection dynamics \citep{Ricci:2002,Ricci:2004}.

The aim of this paper is to study the effect of a background $\mathrm{O^+}$ population on the evolution of magnetic reconnection using realistic parameters. A current sheet composed of $\mathrm{H^+}$ only, and a mixed background of $\mathrm{O^+}$ and $\mathrm{H^+}$ , to mimic the lobe population of the magnetotail, are considered. The results are compared with identical simulations but with only $\mathrm{H^+}$  plasma background. After an overall study of magnetic reconnection evolution, an in depth analysis of the flow pattern and on the different dynamics of background $\mathrm{O^+}$ and $\mathrm{H^+}$ particles is performed to study the role of $\mathrm{O^+}$ in magnetic reconnection.

The paper is organized as follows. The second section presents the simulation parameters and the initial set-up. Three main results are reported in the third section. First, the overall magnetic reconnection evolution and the electromagnetic field at representative times and the different acceleration mechanisms acting on $\mathrm{H^+}$ and $\mathrm{O^+}$ ions are presented. Second, the density evolution of the two species is studied and the initial $\mathrm{H^+}$ $\mathrm{O^+}$ separation in the outflow region is shown. Third, the effect of a guide field on magnetic reconnection with $\mathrm{O^+}$ and the $\mathrm{H^+}$ dynamics is analyzed. The fourth section concludes the paper summarizing the results and discussing the validity of the presented simulations.

\section{Simulation approach and physical system}
Two and half dimensional (two spatial components and three components for the electromagnetic fields and particles velocities) Particle-in-Cell simulations have been completed on the X-Y plane with X along Earth-Sun direction in the Earth's magnetosphere, and with Y pointing to the north-south direction. The X and Y coordinates correspond to the -X and Z coordinates respectively in the GSM system, while in our coordinate system Z is out-of-plane ignorable coordinate. An antiparallel magnetic field, vanishing in the mid-line of the simulation domain ($y = L_y/2$), is initialized in the X direction as:
 \begin{equation}
 B_x=B_0\tanh(\frac{y - L_y/2}{\lambda})
\end{equation}
where  $\lambda$ is the current sheet thickness, and $B_0$ is the lobe magnetic field. 
The number density has the following profile:
\begin{equation}
 n(y) = n_0 \cosh^{-2}(\frac{y - L_y/2}{\lambda})
\end{equation}
where $n_0$ is the peak particle density at the center of the system. The simulation box is $40 d_{H}$ ($L_x$) and $20d_{H}$ ($L_y$) long in the X and Y directions, where $d_{H}$ is the $\mathrm{H^+}$ skin depth based on the peak density $n_0$ of the Harris sheet with $\lambda = 0.5 d_{H}$. The particles forming the current sheet are electrons and only $\mathrm{H^+}$ ions. They are initialized with a Maxwellian distribution with a velocity drift that satisfies the Harris equilibrium force balance. The electron temperature is $(T_e/(m_e c^2))^{1/2} = v_{the}/c = 0.045$, where $c$ is the speed of light in vacuum, $m_e$ and $v_{the}$ are the electron mass and thermal velocity. The $\mathrm{H^+}$ ions have five times the electron temperature, $T_{H} = 5 T_e$. In addition, a tenuous background plasma with $\mathrm{H^+}$  and $\mathrm{O^+}$ densities $n_{bH} = n_{bO} = 0.05 n_0$ is added to mimic the lobe plasma. This background plasma is uniformly distributed in the simulation box; it is not drifting and has the same temperature of the current sheet plasma $T_{H} = T_{O}$. The simulation time step is $\omega_{pH}\Delta t = 0.06$, where $\omega_{pH} =\sqrt{4\pi n_0 e^2/m_H}$ is the $\mathrm{H^+}$ plasma frequency, $m_H$ is the $\mathrm{H^+}$ mass and $e$ is the elementary charge. In this simulation set-up, the ratio $\omega_{pH}$ to $\Omega_{cH} = B_0e/m_H$ ($\mathrm{H^+}$ gyro-frequency), that is equal to the ratio of $c$ to $V_{AH}$, the $\mathrm{H^+}$ Alfv\'en velocity calculated with $B= B_0$ and $n = n_0$, results 275. This ratio is comparable to the typical value of 300, observed in the Earth's magnetotail\citep{Lapenta:2010}. The grid is composed of $1024\times512$ cells. In total 411,041,792 computational particles with physical mass ratios $m_{H}/m_{e} = 1836$ and $m_{O}/m_{H} = 16$ are used in all the reported simulations. A localized perturbation around the X point is applied to initiate magnetic reconnection without a boundary driving avoiding the creation of spurious compressive waves in the simulation domain \citep{Lapenta:2010}:
 \begin{equation}\label{perturb}
\delta A_z=A_{z0} \cos(2\pi (x-L_x/2)/L_\Delta)cos(\pi (y-L_y/2)/L_\Delta)e^{-((x-L_x/2)^2+(y-L_y/2)^2)/\sigma^2},
 \end{equation}
with $L_\Delta=10 \sigma$ and $\sigma=0.5d_{H}$. The boundary conditions are periodic in the X direction, while conducting and reflecting boundary conditions are imposed  in the Y direction for the electromagnetic field and particles.
A simulation with only $\mathrm{H^+}$ ions in the background plasma has been completed to compare the results and determine the effect of the $\mathrm{O^+}$ species in the magnetic reconnection. The companion simulation has identical parameters but starts from a background population, that is composed only of  $\mathrm{H^+}$ with initial density $n_{bH} = 0.1 n_0$. Simulations are completed using the parallel Particle-in-Cell code {\em iPIC3D} \citep{Markidis:2010}, that uses the moment implicit Particle-in-Cell scheme and allows large time step and grid spacing still retaining the numerical stability. 

\section{Simulation results}
The reconnected flux $\Delta \Psi$ is calculated as the difference of the magnetic flux between the X and O points and it is studied to characterize the speed of magnetic reconnection. Its evolution is shown for the two simulations with a mixed population of $\mathrm{O^+}$ and $\mathrm{H^+}$ ions and with only the $\mathrm{H^+}$ species in panel a) of Figure 1. A slow growth of the reconnected flux is visible until $\Omega_{cH} t \approx 5$, then the fast phase starts. The $\mathrm{O^+}$ presence slightly affects the evolution of magnetic reconnection: the reconnected flux is approximately the same for both simulations. The reconnection rate is calculated as the derivative in time of the reconnected flux computed every $1.1 \Omega_{cH}^{-1}$. It is normalized to the $B_0V_{AH}/c$, and shown in panel b) of Figure 1. It increases until time  $\Omega_{cH} t \approx 10$, and reaches the peak value approximately at time $\Omega_{cH} t \approx 12$. When the two simulations with different ion background are compared, the reconnection rate peaks differ by 0.034 $B_0V_{AH}/c$, approximately 10 \% of the maximum reconnection rate. 

\begin{figure*}
\label{ReconFlux}
\noindent
\center
\includegraphics[width=0.8\textwidth,angle=0]{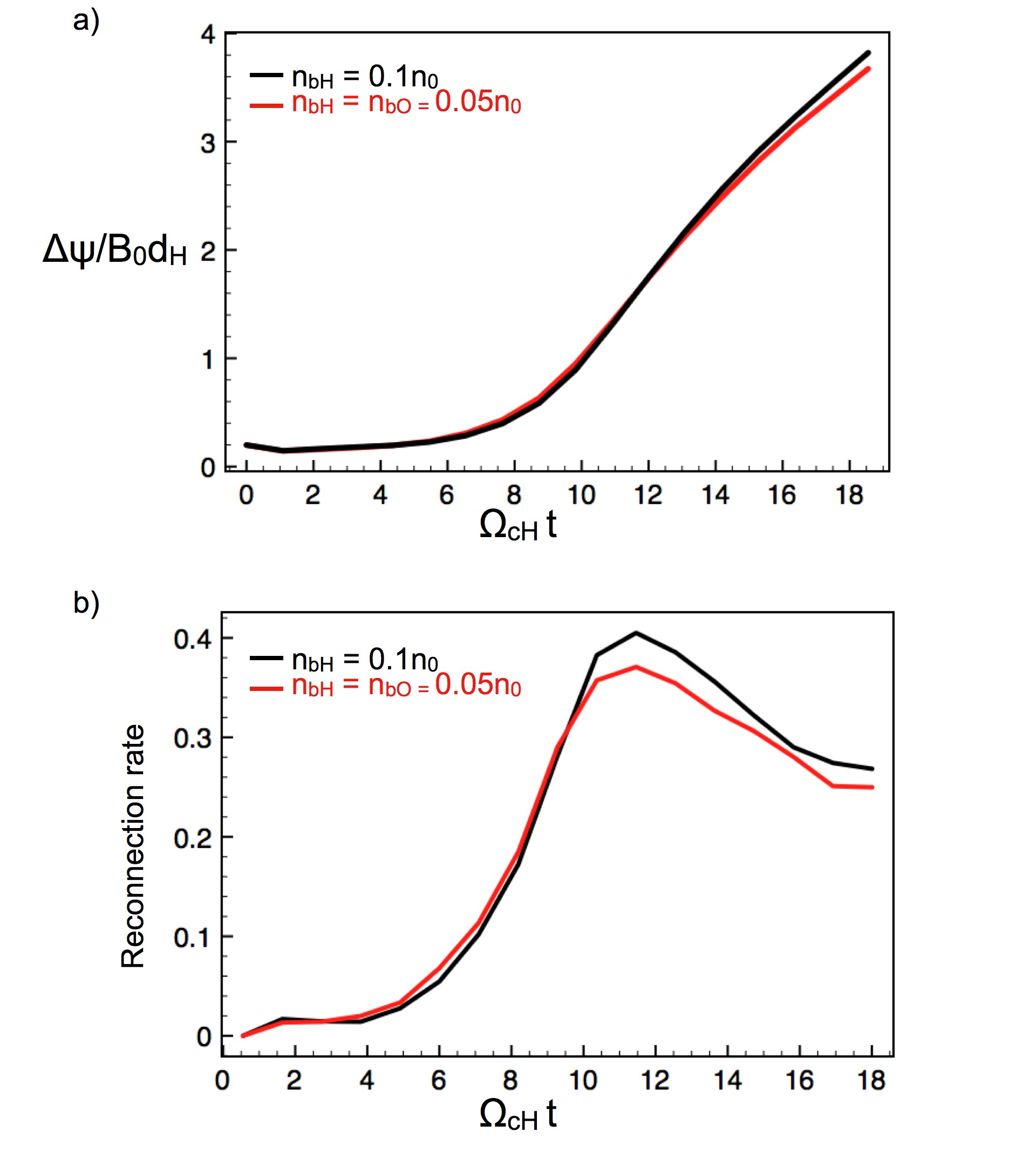} 
\caption{Evolution of the reconnected flux $\Delta \Psi$, normalized to $B_0 d_{H}$, and reconnection rate, normalized to the $B_0V_{AH}/c$, for a simulation with only a background $\mathrm{H^+}$ population with $n_{bH} = 0.1n_0$ in black and for a simulation with mixed background population of $\mathrm{H^+}$ and $\mathrm{O^+}$ ions with $n_{bH} =  n_{bO} = 0.05n_0$ in red.}
\end{figure*}

The pseudocolor plots of different components of the electric field at time $\Omega_{cH} t= 10.9$, when the reconnection rate is maximum, are presented in three panels of Figure 2. The reconnection electric field, $E_z$ in the chosen reference system, is shown in panel a). The peak value of the reconnection field is $8 \times 10^{-6} (\omega_{pH} m_{H} c)/e$. The in-plane electric fields are localized along the reconnection fronts, and are characterized by higher values around $2-4 \times 10^{-5} (\omega_{pH} m_{H}c)/e$, as visible in the pseudocolor plots in panels b) and c) of Figure 2.

\begin{figure*}
\label{EfieldsRecon}
\noindent
\center
\includegraphics[width=.70\textwidth,angle=0]{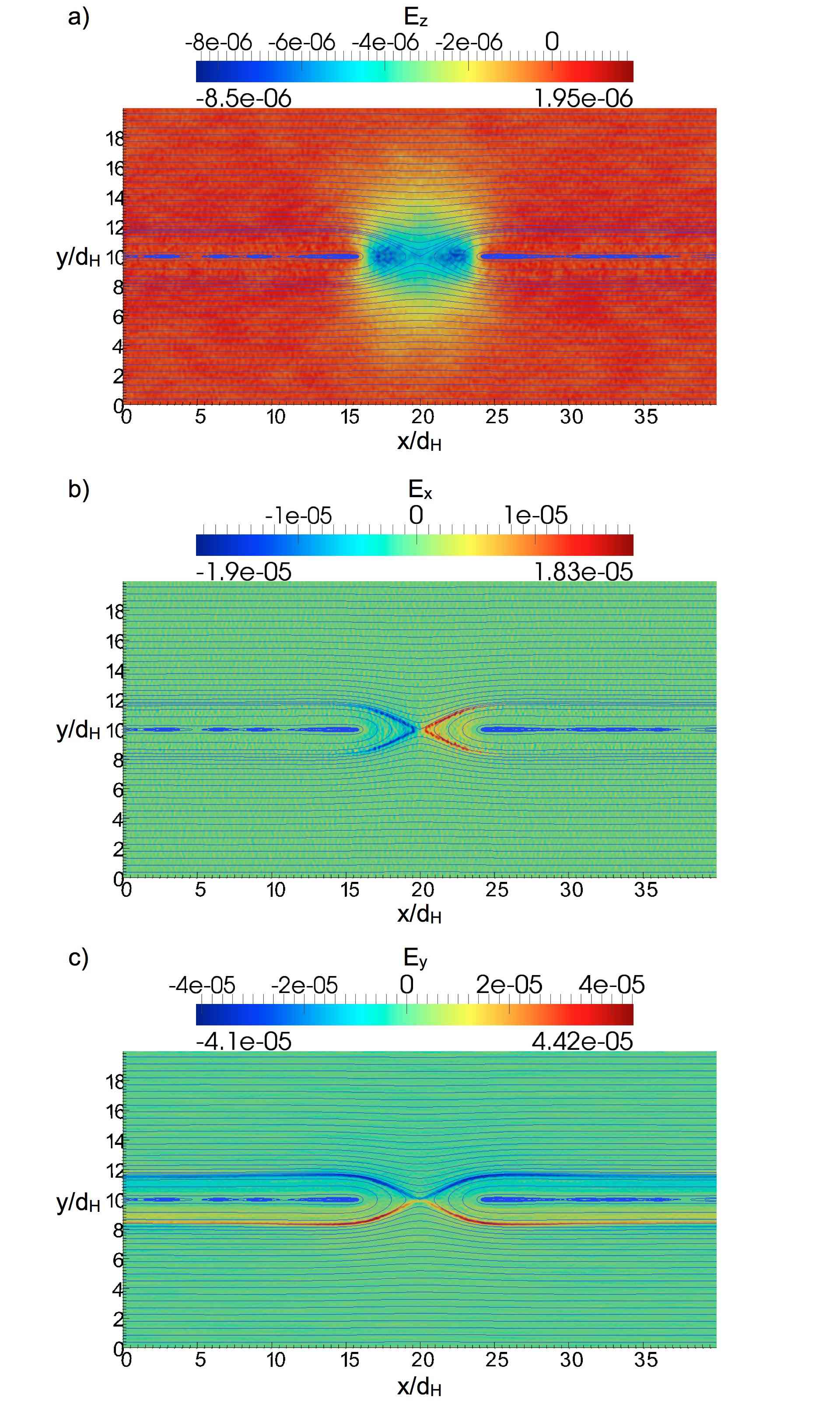} 
\caption{Reconnection electric field $E_z$, electric field X and Y components in units of $(\omega_{pH} m_{H} c)/e$ with superimposed magnetic field lines in blue at time $\Omega_{cH} t= 10.9$.}
\end{figure*}

A comparison of the quadrupolar out-of-plane component of magnetic field, signature of Hall reconnection, at time $\Omega_{cH} t= 16.4$ is shown in Figure 3. The $B_z$ component for the simulation with and without the $\mathrm{O^+}$ background species are presented respectively in panels a) and b). Clearly, the typical quadrupolar structure reaches the same peak values (approximately $0.0009 \omega_{pH} m_{H} c/e = 0.25 B_0$), but it is broader in the case of the simulation with mixed $\mathrm{H^+}$, $\mathrm{O^+}$ background populations. In addition, Figure 4 presents the line profile of the out-of-plane component of the magnetic field $B_z$ along $x = 8d_{H}$ to compare the peak values and the thickness of the quadrupolar structure. In proximity of the neutral line ($y=10d_{H}$), two small $B_z$ peaks reveal the electron scales. Moreover, the Hall magnetic field profile in presence of $\mathrm{O^+}$ (red line in Figure 4) reaches peak values in a relatively small region, starting from the neutral line and ending at $4d_{H}$ from it, and then slowly decreases over an extended area reaching the boundary. Thus, the $B_z$ profile can be divided in two parts: the sharp peaks representing the "core" Hall magnetic field, and a slowly decreasing $B_z$ representing the "tail" Hall magnetic field. This two-scale nature of the Hall magnetic field reflects the presence of the two $\mathrm{H^+}$ and $\mathrm{O^+}$ dissipation layers, and of light and heavy whistler scales \citep{Shay:2004}. 

\begin{figure*}
\label{HallField}
\noindent
\center
\includegraphics[width=1.0\textwidth,angle=0]{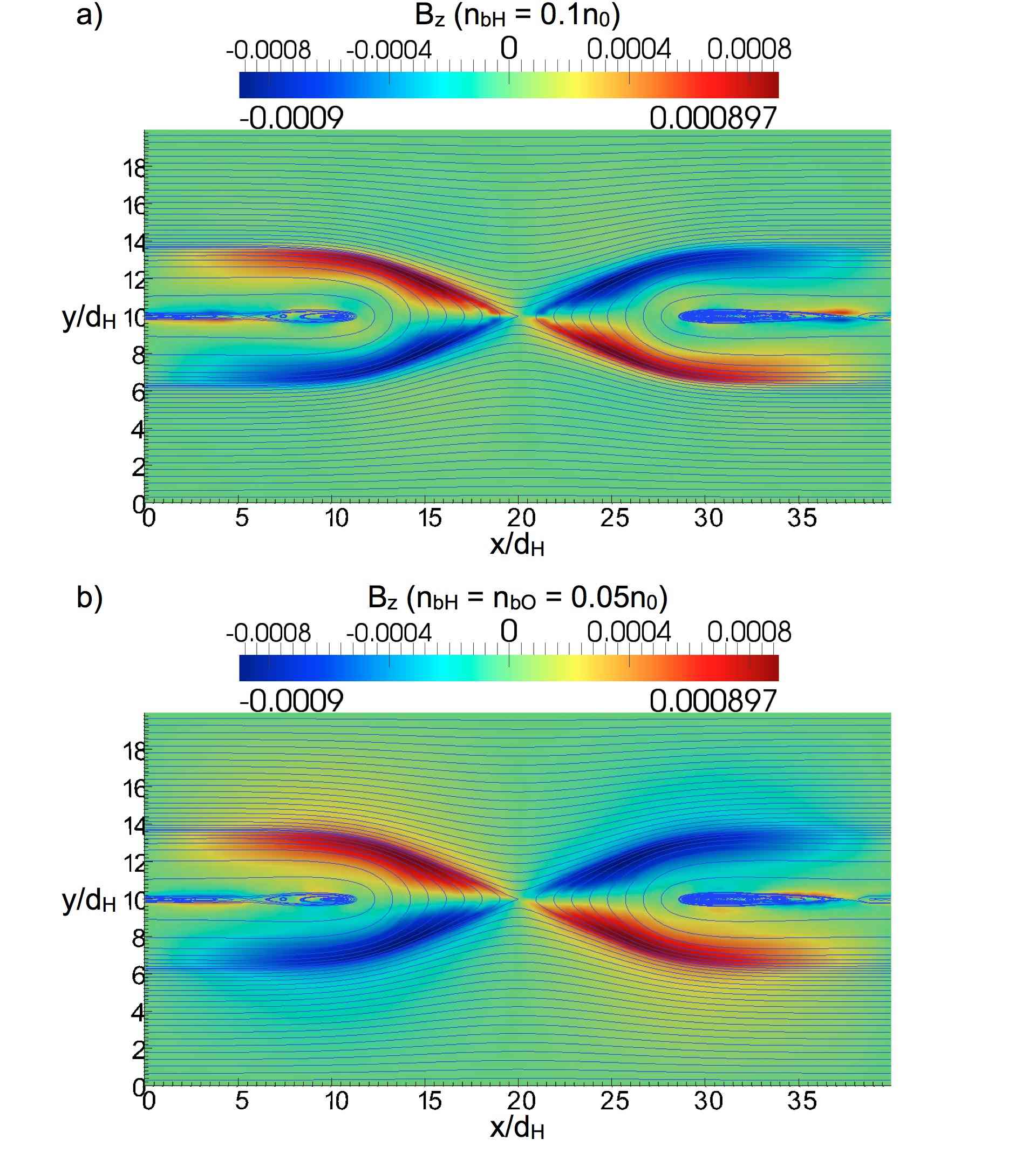} 
\caption{Out-of-plane magnetic field component $B_z$, in units of $(\omega_{pH} m_{H} c)/e$, at time $\Omega_{cH} t= 16.4$  for a simulation with only a background $\mathrm{H^+}$ population with $n_{bH} = 0.1n_0$ in panel a) and for a simulation with mixed background population of $\mathrm{H^+}$ and $\mathrm{O^+}$ ions with $n_{bH} =  n_{bO} = 0.05n_0$ in panel b). The quadrupolar structure reaches the same peak values (approximately $0.0009 \omega_{pH} m_{H} c/e = 0.25 B_0$) in both cases, but it is broader in the case of the simulation with mixed $\mathrm{H^+}$, $\mathrm{O^+}$ background populations.}
\end{figure*}

\begin{figure*}
\label{HallField_2}
\noindent
\center
\includegraphics[width=1.0\textwidth,angle=0]{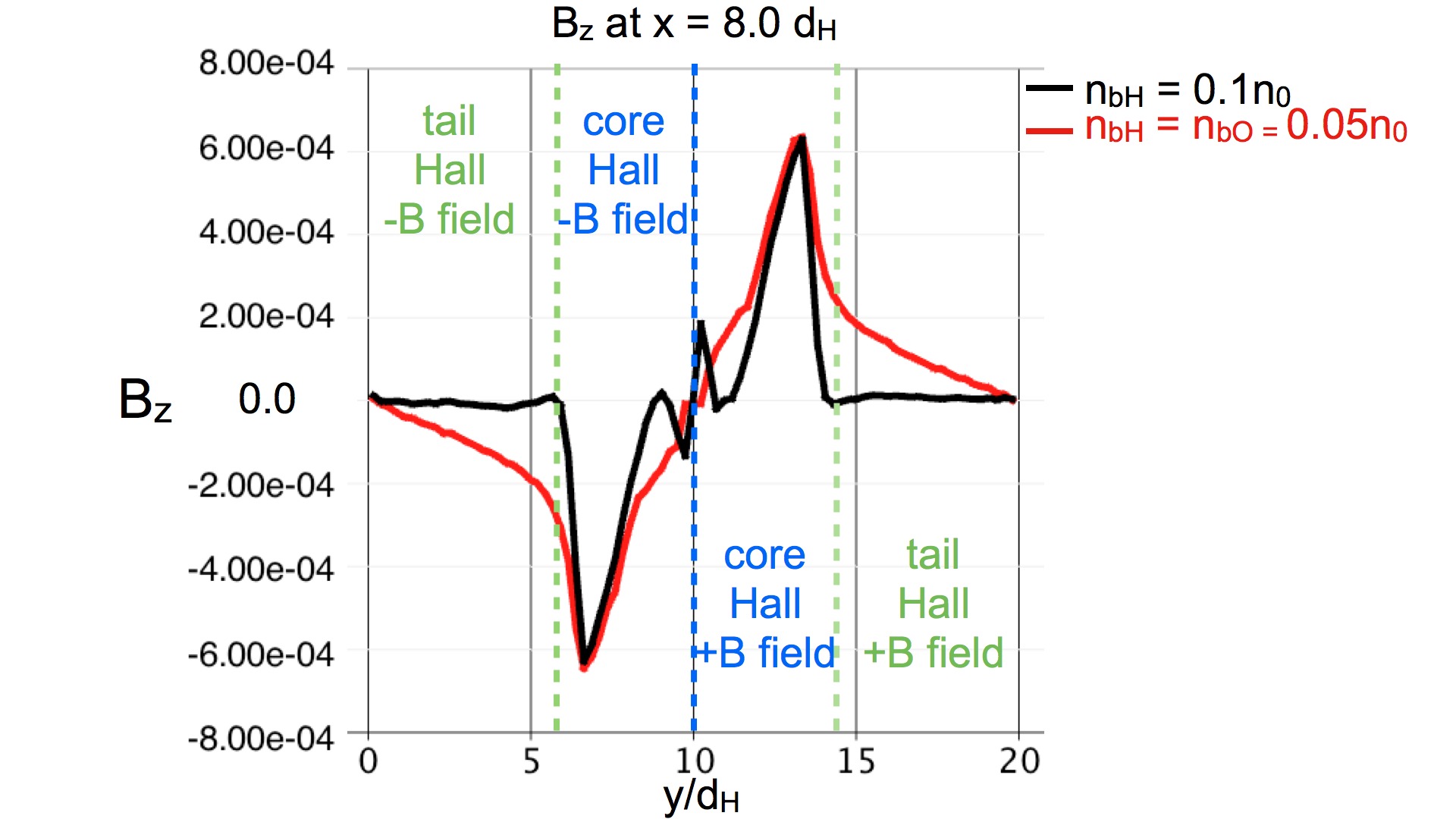} 
\caption{Hall magnetic field $B_z$ profile along the $x=8.0d_{H}$ line for the simulation with mixed $\mathrm{O^+}$ and $\mathrm{H^+}$ (red line) and only $\mathrm{H^+}$ (black line) background. The small peaks in proximity of the neutral line ($y=10d_{H}$) reveal the electron scales. In presence of an $\mathrm{O^+}$ population, the Hall magnetic field $B_z$ profile (red line) presents a two-scale structure: sharp $B_z$ peaks (core Hall magnetic field) and slowly decreasing $B_z$ (tail Hall magnetic field). The $B_z$ peak values are approximately $0.0006 \omega_{pH} m_{H} c/e = 0.16 B_0$.}
\end{figure*}

The magnitude of the average magnetic force divided by the charge $e$ ($\mathbf{v _s}\times \mathbf{B}/c$), where $v_s$ is the average velocity for species $s$, is plotted in Figure 5 to compare the effectiveness of the electric and magnetic fields particle acceleration on $\mathrm{H^+}$ and $\mathrm{O^+}$ species. It is clear from panel a) of Figure 5 that the magnetic force acting on the $\mathrm{H^+}$ background population is maximum along the separatrices and in a shell around the X point. The immediate region around the X point has vanishing magnetic field, and the $\mathrm{H^+}$ ions in it are demagnetized. The maximum intensity value is $8.42\times10^{-6} \omega_{pH} m_{H} c/e$. The average magnetic force acting on $\mathrm{O^+}$ particles is shown in panel b), and its maximum values is four times less than the $\mathrm{H^+}$ one, $2.14\times10^{-6} \omega_{pH} m_{H} c/e$.  

\begin{figure*}
\label{MagneticForce}
\noindent
\center
\includegraphics[width=1.0\textwidth,angle=0]{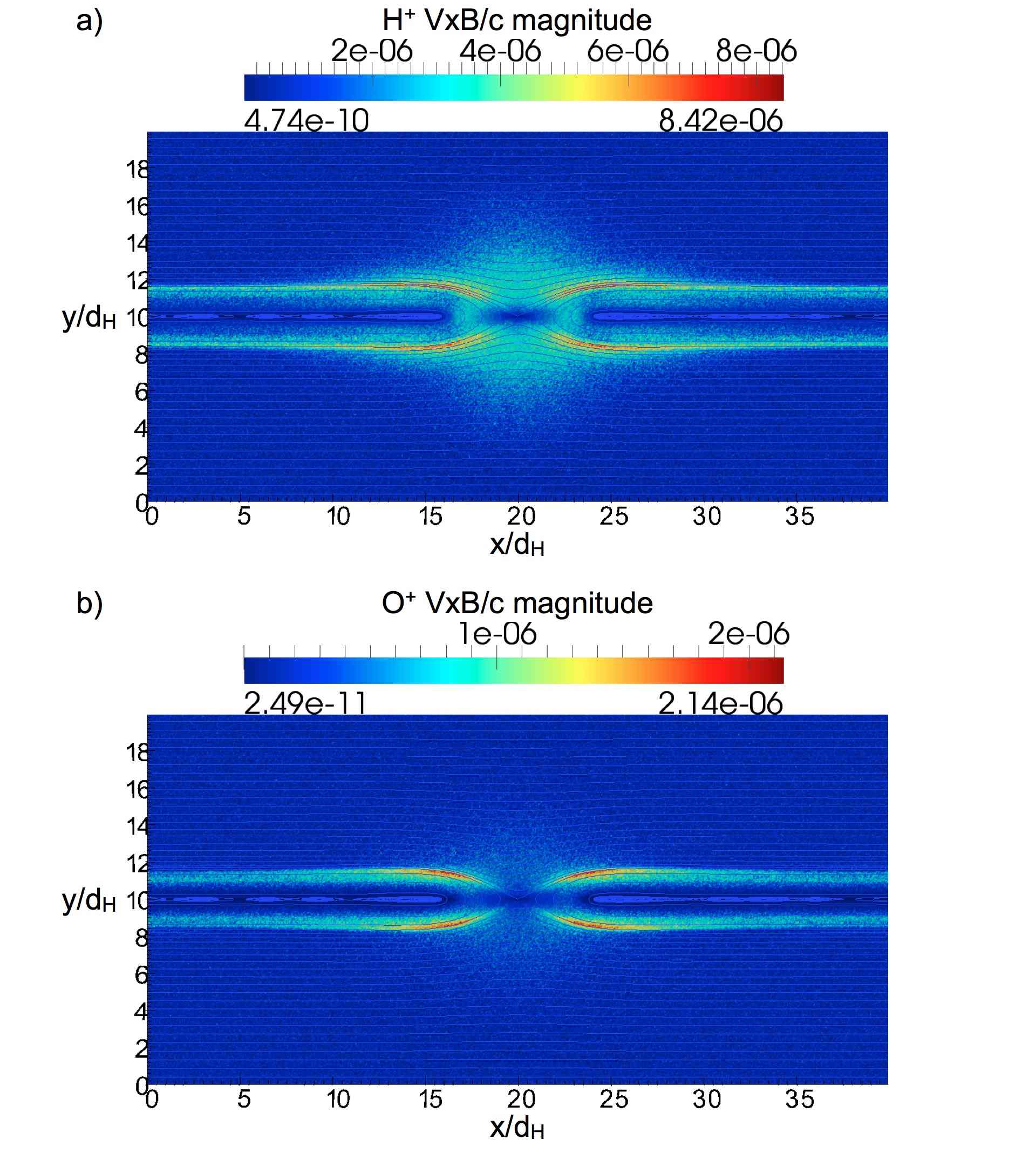} \\
\caption{$\mathbf{v} \times \mathbf{B}/c$ {\em magnetic} force magnitude for the $\mathrm{H^+}$ and $\mathrm{O^+}$ background populations with superimposed magnetic field lines in blue at time $\Omega_{cH} t= 10.9$.}
\end{figure*}

\subsection{Acceleration mechanisms}
It is useful to study the flow patterns and velocities for the two ion background species to understand different acceleration mechanisms. Figure 6 shows the in-plane (X-Y plane) average velocity for the $\mathrm{H^+}$ and $\mathrm{O^+}$ background populations at time $\Omega_{cH} t= 18.5$. At this time, $\mathrm{H^+}$ background ions have high velocity mainly in the outflow region. The $\mathrm{H^+}$ peak velocity reaches the $0.00461c$ value ($1.27 V_{AH}$). On the other hand, the $\mathrm{O^+}$  species is characterized by a rather different velocity pattern as clear in panel b): its flow is localized in the inflow region, and it is normal to the reconnection separatrices. The peak velocity is $0.001405c$ ($0.39 V_{AH}$), approximately three time less than background $\mathrm{H^+}$ velocity. In addition, a small part of the $\mathrm{O^+}$ population in the X point region participates in the outflow with an approximately $0.0005c$ velocity ($0.14  V_{AH}$).

\begin{figure*}
\label{QuiverPlot}
\noindent
\center
\includegraphics[width=1.0\textwidth,angle=0]{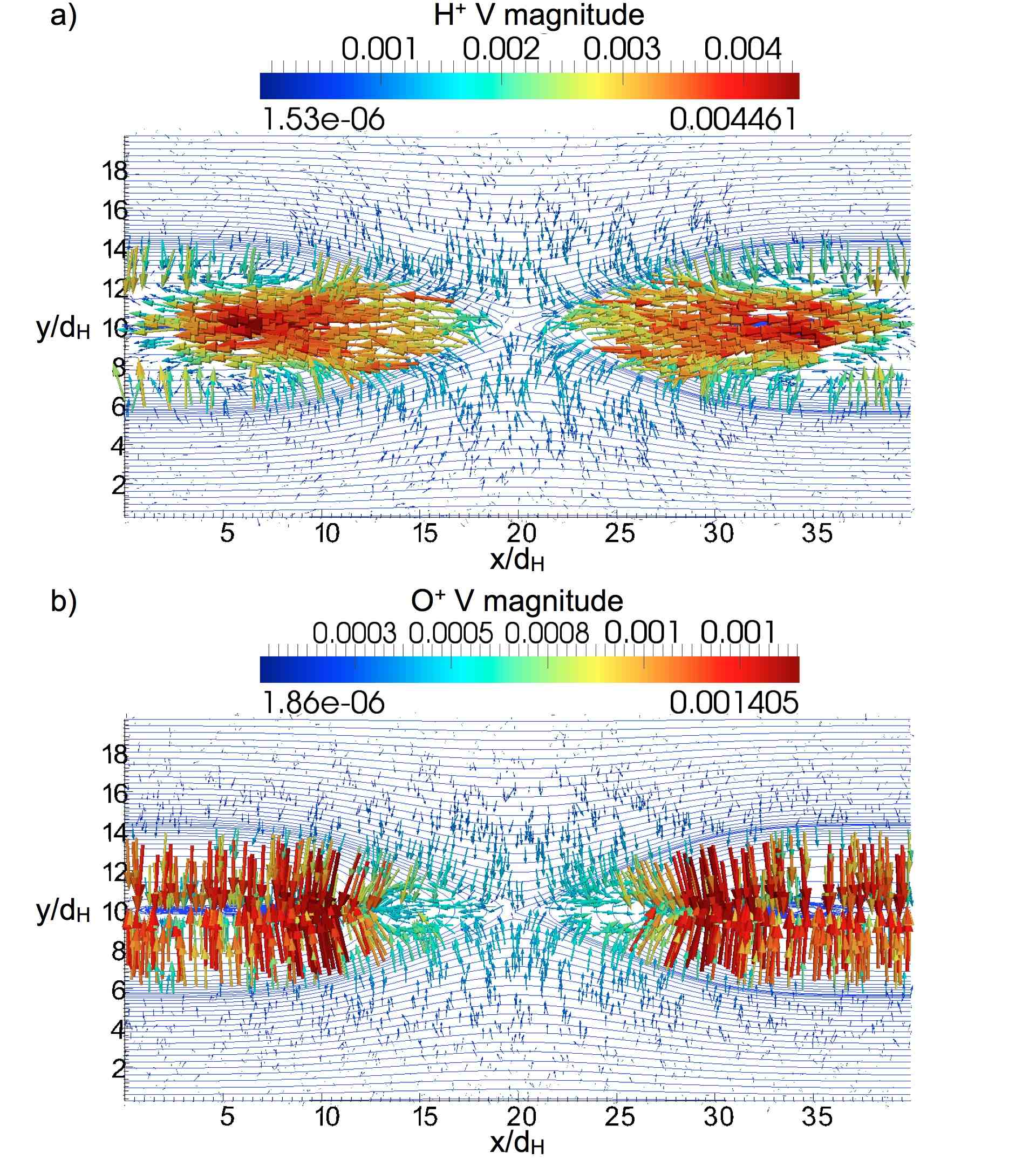}
\caption{In-plane average velocity for $\mathrm{H^+}$ and $\mathrm{O^+}$ background species at time $\Omega_{cH} t= 18.5$. The arrow colors and lengths indicate the velocity magnitude, normalized to $c$. The maximum velocities $0.00461c$ and $0.001405c$ correspond to $1.27 V_{AH}$ and $0.39 V_{AH}$.}
\end{figure*}

A plot of the streamlines at time $\Omega_{cH} t= 18.5$ is shown in Figure 7. The streamlines are calculated as tangents to the average velocities in the X-Y plane at a given time. In panel a) the $\mathrm{H^+}$ flow, represented by black lines, moves from the inflow region towards the X point and it is deflected in the outflow region. The thickness of the outflow is approximately $3d_{H}$. Instead, the $\mathrm{O^+}$ ions follow a rather different flow pattern, which is represented with red lines in panel b). A large part of the $\mathrm{O^+}$ ions moves from the outflow and is deflected perpendicularly to the separatrices in proximity of the reconnection fronts. A small part of the $\mathrm{O^+}$ population is deflected by a $90^{\circ}$ angle around the X point. The simulations show that the acceleration acting on $\mathrm{O^+}$ ions could be explained by the mechanism proposed in Refs. \citep{Drake:2009a,Drake:2009b}: $\mathrm{O^+}$ particles behave like pick-up ions and are accelerated by magnetic reconnection exhausts in the boundary layer between the upstream and downstream plasmas.

\begin{figure*}
\label{StreamingLines}
\noindent
\center
\includegraphics[width=1.0\textwidth,angle=0]{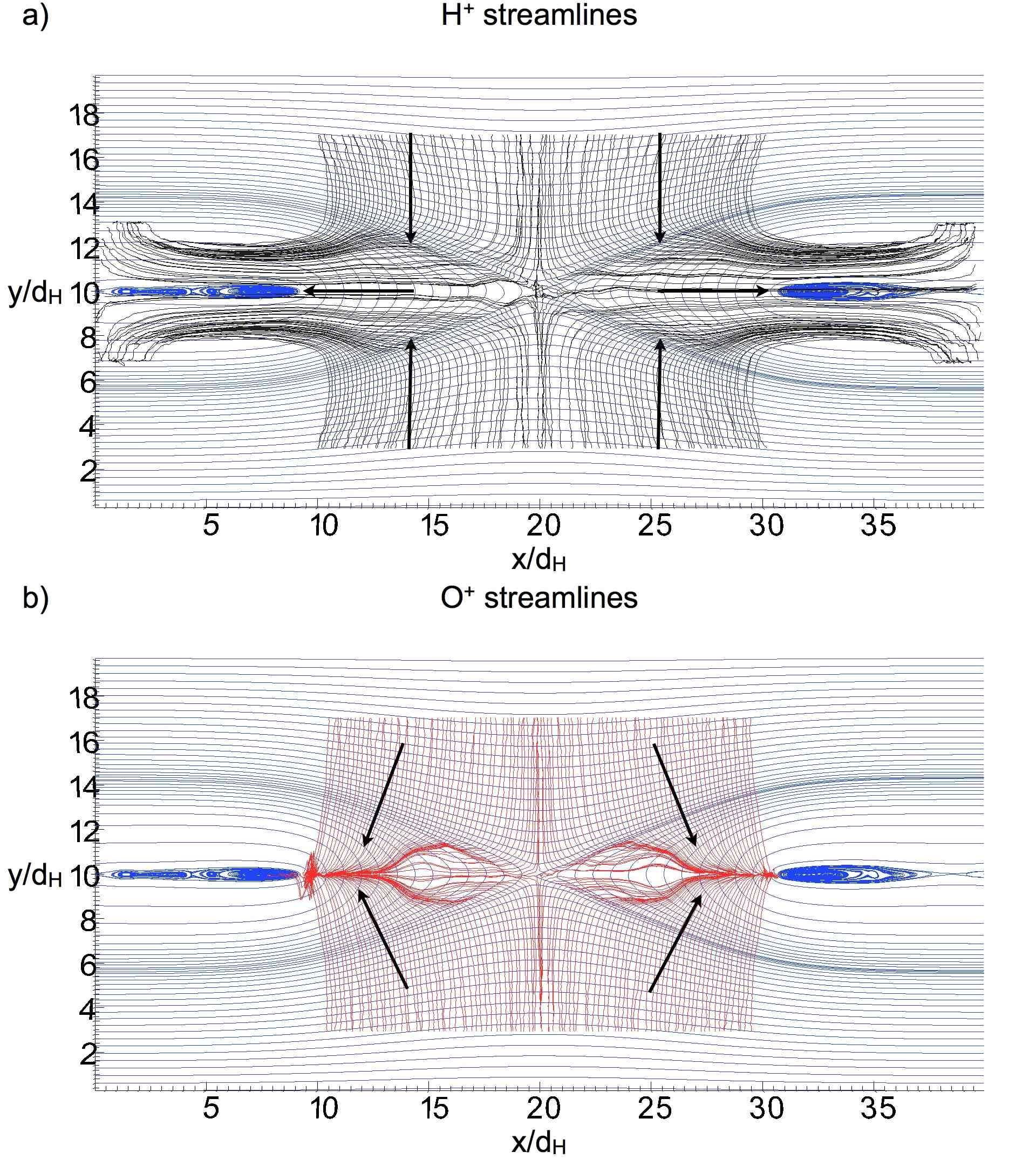} 
\caption{In-plane streamlines for the background $\mathrm{H^+}$ and $\mathrm{O^+}$ populations in black and red with superimposed magnetic lines in blue at time $\Omega_{cH} t= 18.5$. The black arrows show the flow direction.}
\end{figure*}

\subsection{Species separation}
The different acceleration and dynamics of background ions have important macroscopic consequences. The $\mathrm{H^+}$ background population is accelerated towards the X point by the $\mathbf{E} \times \mathbf{B}$ drift \citep{Pritchett:2001} and expelled in the outflow region. The $\mathrm{H^+}$ species remains concentrated on the dipolarization fronts. Instead, the $\mathrm{O^+}$ component does not effectively experience a magnetic drift and responds mainly to the normal electric field in correspondence of separatrices. These different acceleration mechanisms determine enhanced density areas of $\mathrm{H^+}$, and $\mathrm{O^+}$ in different regions. Figure 8 shows a pseudocolor plot of the density, normalized to $n_0$, at time $\Omega_{cH} t= 18.5$ for the $\mathrm{O^+}$ and $\mathrm{H^+}$ background populations. The initial value density for both species is $n_{bH} =  n_{bO} = 0.05n_0$. A comparison between the two panels of Figure 8 reveals that peak density is higher in the case of $\mathrm{H^+}$, that reaches a value of approximately $0.2n_0$, while the peak density for $\mathrm{O^+}$ population is $0.15n_0$. Moreover, the two background species concentrate in different parts of the system. This is clear from panel a) of Figure 9, a pseudocolor plot of the peak densities at $\Omega_{cH} t= 18.5$. The peak density plots show the regions with density values between the maximum and 70\% of the maximum. The background $\mathrm{H^+}$, $\mathrm{O^+}$ ions and electrons appear respectively as black, red and green areas. In particular, the $\mathrm{H^+}$ and electron peak density is localized on the dipolarization fronts. Instead, the $\mathrm{O^+}$ population is slower and follows the $\mathrm{H^+}$ species in a more extended region from the tail of the $\mathrm{H^+}$ peak density to the X point. Panel b) of Figure 9 shows the line profile of the three species densities along the $y = L_y/2$ line using the same color scheme of panel a). Figure 10 presents the $\mathrm{H^+}$ and $\mathrm{O^+}$ peak densities at three different times, showing the progressive separation of the two ion species in the outflow region at the initial stage of magnetic reconnection.

\begin{figure*}
\label{Densities}
\noindent
\center
\includegraphics[width=1.0\textwidth,angle=0]{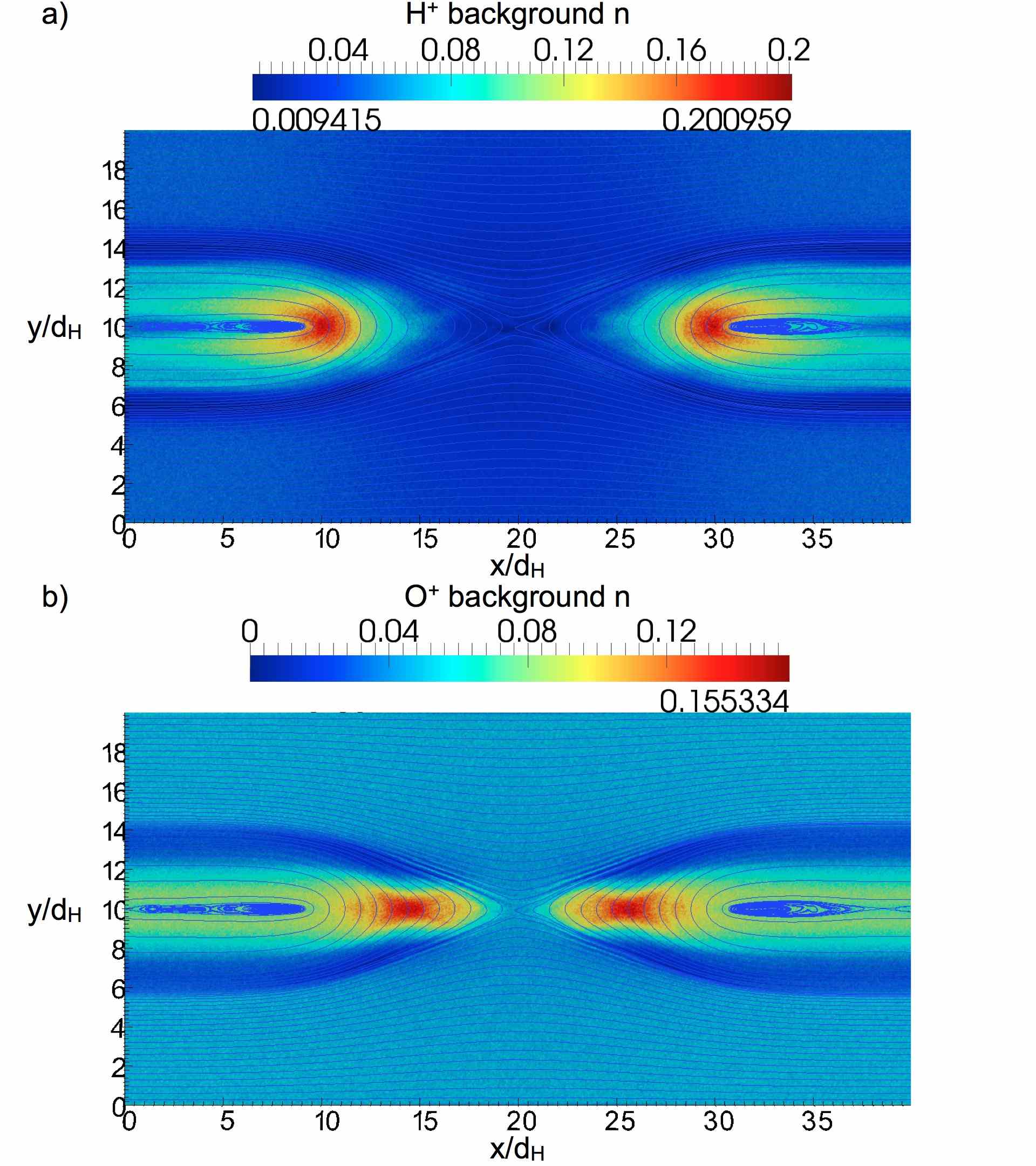} 
\caption{Densities, normalized to $n_0$,  for the background $\mathrm{H^+}$ and $\mathrm{O^+}$ populations at time $\Omega_{cH} t= 18.5$ with superimposed magnetic field lines in blue.}
\end{figure*}

\begin{figure*}
\label{Separation}
\noindent
\center
\includegraphics[width=1.0\textwidth,angle=0]{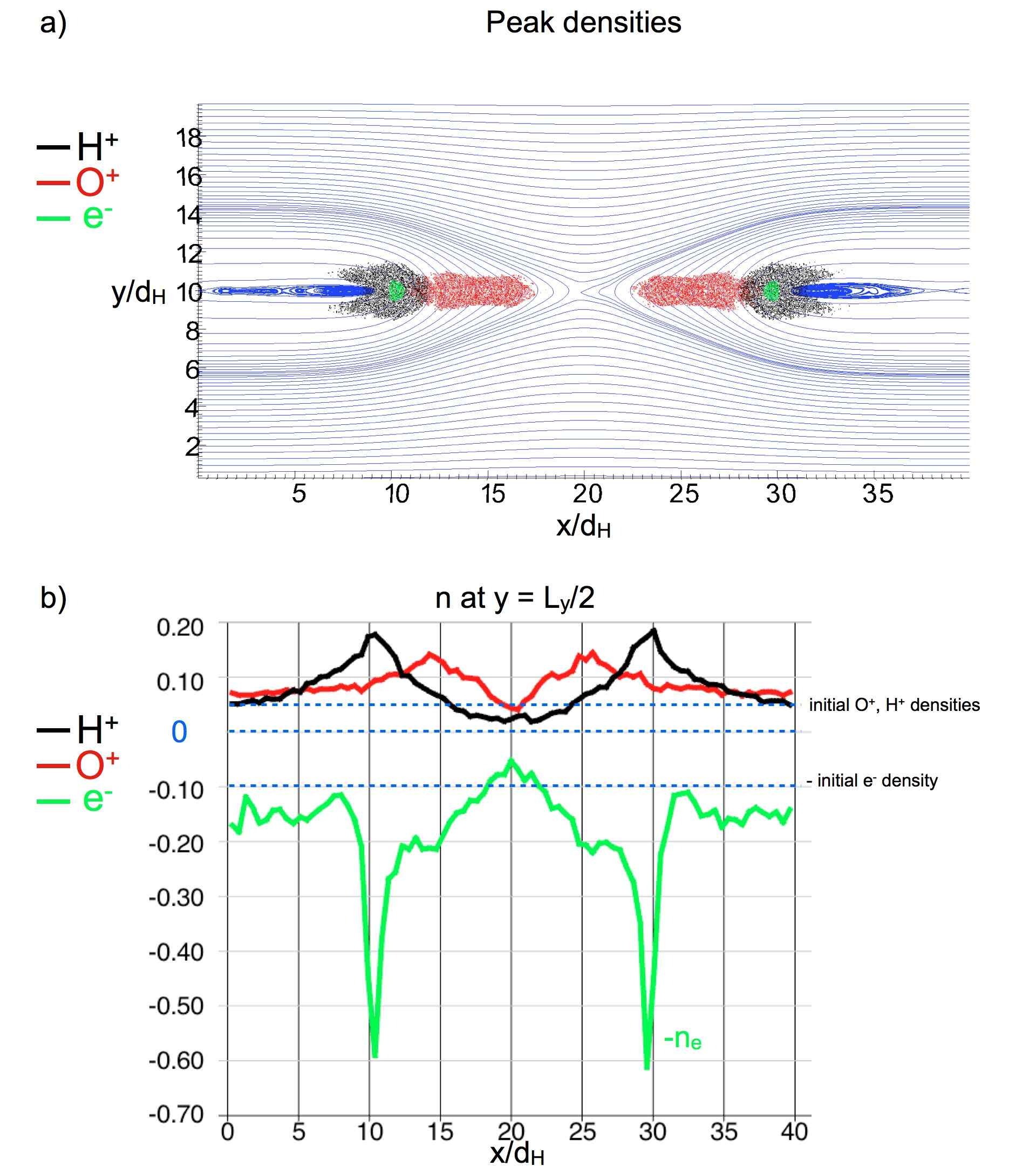} 
\caption{Peak densities regions for $\mathrm{H^+}$ and $\mathrm{O^+}$ and electron background populations in black, red and green respectively at time $\Omega_{cH} t= 18.5$ with superimposed magnetic field lines in blue in panel a). In panel b) the density profiles for the three background species along the line $y= L_y/2$ at time $\Omega_{cH} t= 18.5$ with the same coloring scheme is shown.}
\end{figure*}

\begin{figure}
\label{PeakDensities2}
\noindent
\center
\includegraphics[width=\columnwidth,angle=0]{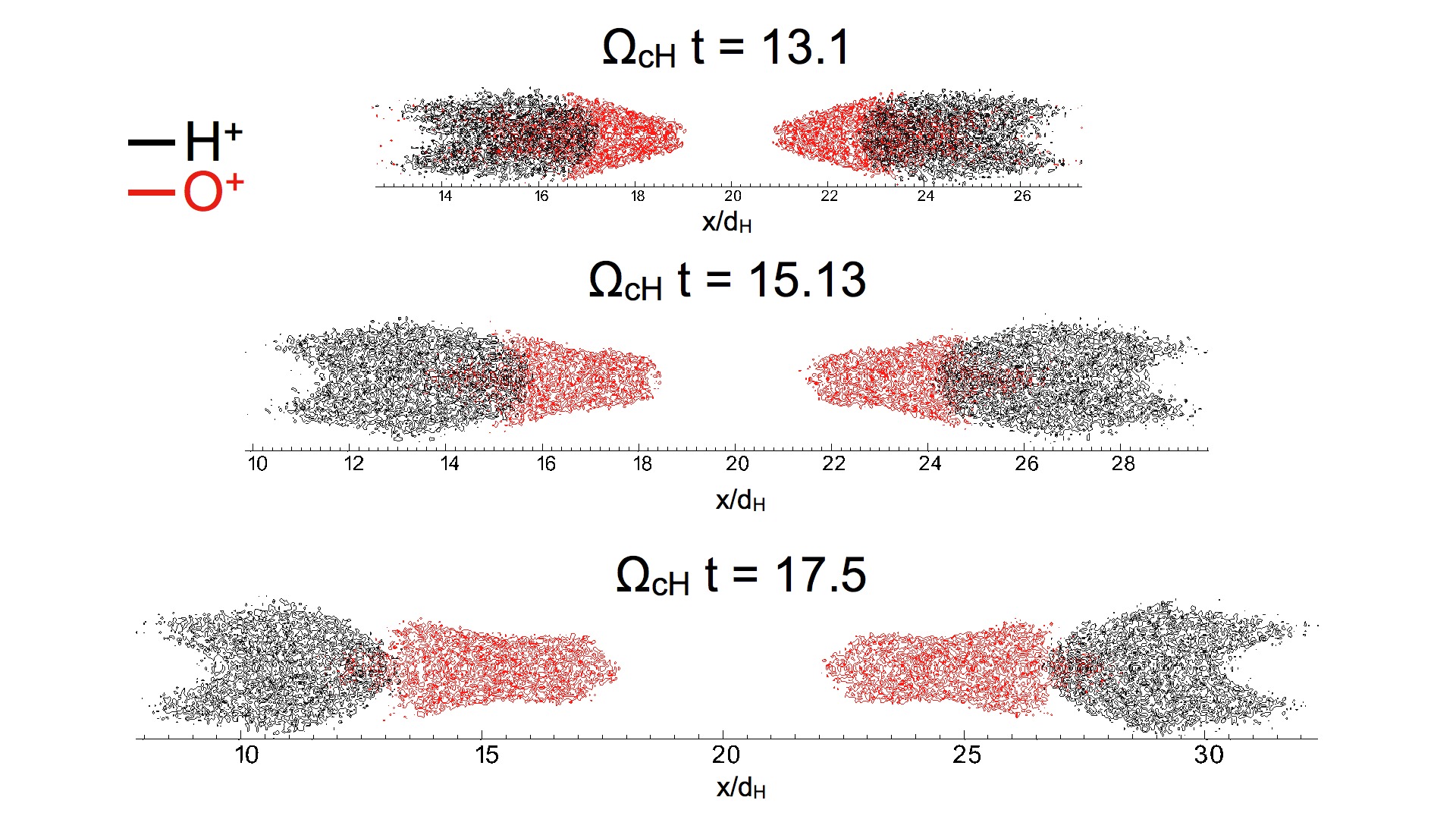}
\caption{Peak densities regions for $\mathrm{H^+}$ and $\mathrm{O^+}$ background populations in black and red colors at three successive times. The $\mathrm{H^+}$ and $\mathrm{O^+}$ populations progressively separate in the outflow region at the initial stage of magnetic reconnection.}
\end{figure}

\subsection{The effect of a guide field}
To mimic the configuration of magnetic reconnection in the magnetopause, it is useful to add an uniform out-of-plane guide field in the initial Harris sheet equilibrium \citep{Ricci:2004,Swisdak:2005,Lapenta:2010,Goldman:2010,Lindstedt:2010}. A guide field, perpendicular to the X-Y plane, $B_g = B_0$ is imposed in the initial magnetic configuration. Figure 11 shows a comparison between the reconnected flux between two simulations with $\mathrm{O^+}$ and with and without the guide field in red solid and dashed lines, and a simulation with only background $\mathrm{H^+}$ and guide field in black line. Both simulations with guide field present a slow phase where the reconnected flux increases slowly, that is followed by a fast phase. The reconnected flux is almost identical for both simulations with and without $\mathrm{O^+}$ in presence of a guide field. By comparing the reconnected flux with and without the guide field, it is clear that a guide field $B_0$ reduces the speed of magnetic reconnection to approximately 32\% in the case of simulations with only $\mathrm{H^+}$ ions. In fact, the reconnection rate peaks for simulations with and without guide field are respectively $0.4B_0V_{AH}/c$ and $0.27B_0V_{AH}/c$. These results are in reasonable agreement with those reported by previous studies \citep{Pritchett:2001,Ricci:2004}.

\begin{figure}
\label{ReconFluxGF}
\noindent
\center
\includegraphics[width=\columnwidth,angle=0]{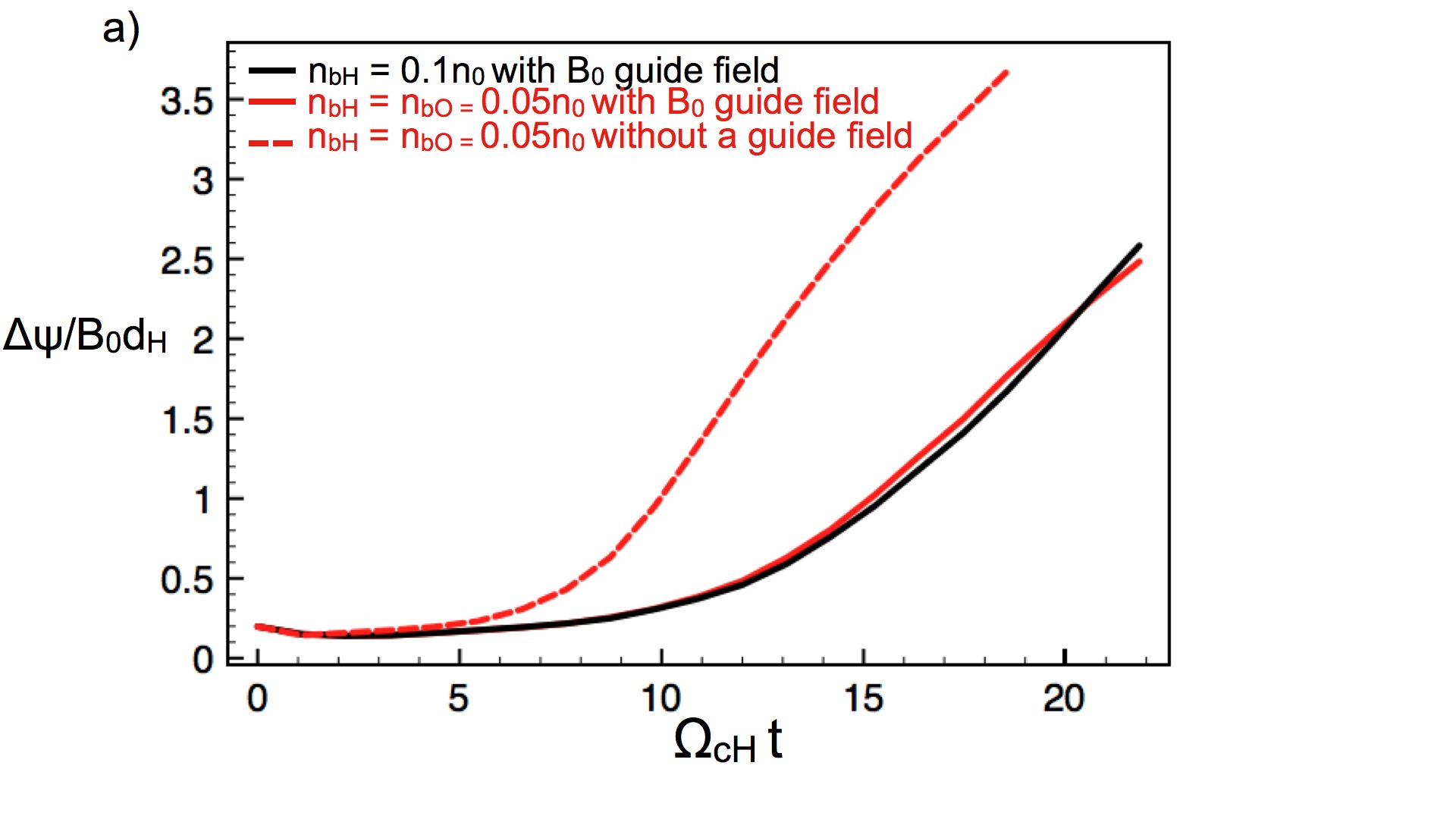}
\caption{Evolution of the reconnected flux $\Delta \Psi$, normalized to $B_0 d_{H}$ for a simulation with mixed background population of $\mathrm{H^+}$ and $\mathrm{O^+}$ ions with $n_{bH} =  n_{bO} = 0.05n_0$, with a guide field in solid red line and without it in dashed red line. The reconnected flux with only background $\mathrm{H^+}$ population and $n_{bH} = 0.1n_0$ is presented in black.}
\end{figure}

Panels a) and b) of Figure 12 show a quiver plot of the average velocity for background $\mathrm{H^+}$ and $\mathrm{O^+}$ populations at time $\Omega_{cH} t= 21.8$. The guide field deviates the background $\mathrm{H^+}$ in the outflow region: the $\mathrm{H^+}$ outflow motion is tilted by the guide field and asymmetric with respect to the X point. The $\mathrm{O^+}$ acceleration is still localized along the separatrices, and normal to them. The peak velocities for $\mathrm{H^+}$ and $\mathrm{O^+}$ with and without guide field are very close in value.

\begin{figure*}
\label{Quiver_GF}
\noindent
\center
\includegraphics[width=1.0\textwidth,angle=0]{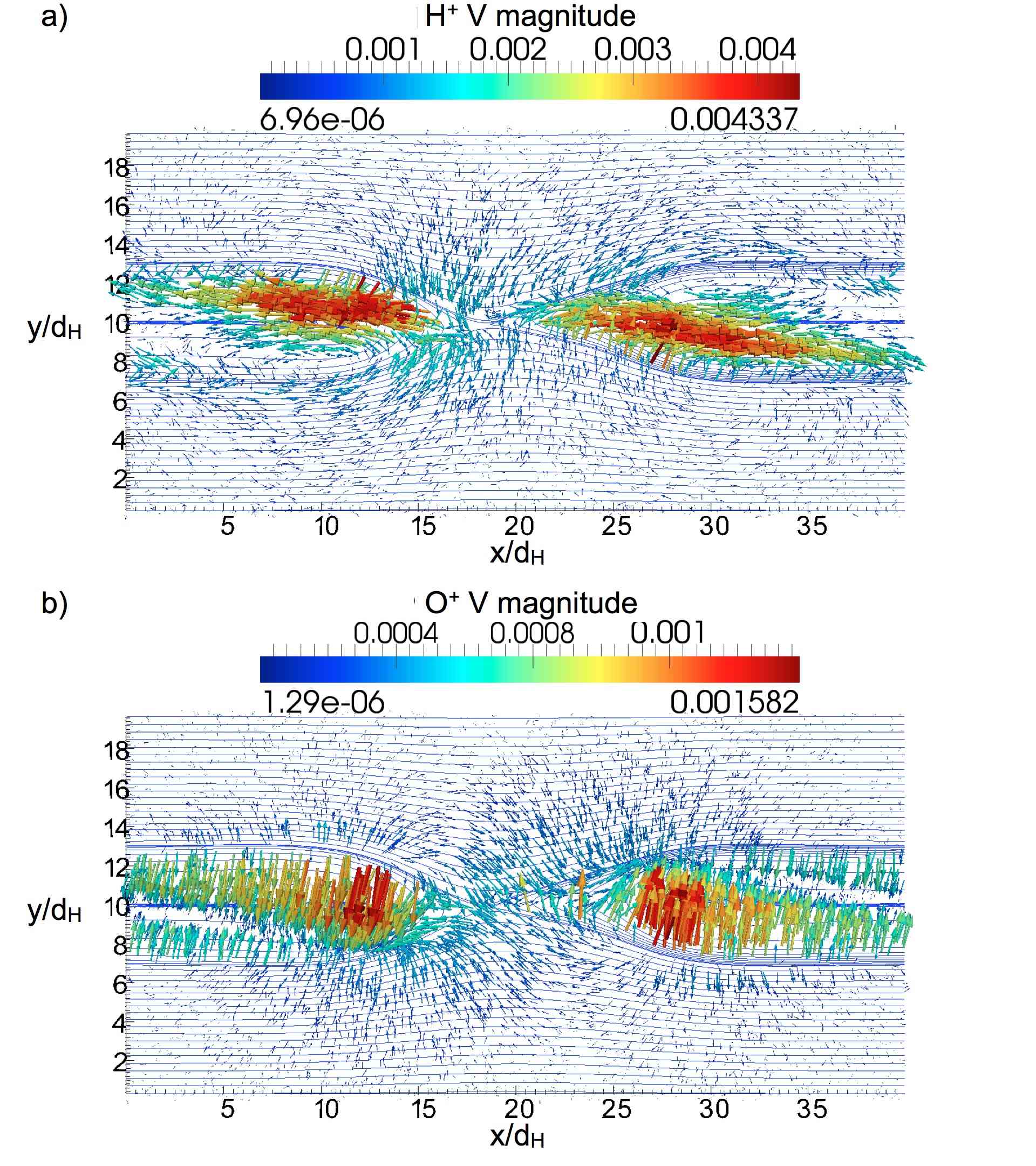} \\
\caption{In-plane average velocity of $\mathrm{H^+}$ (panel a) and $\mathrm{O^+}$ (panel b) background species at time $\Omega_{cH} t= 21.8$ in presence of a guide field. The magnitude of average velocity is normalized to $c$. The maximum velocities $0.004337c$ and $0.001582c$ correspond to $1.19 V_{AH}$ and $0.43 V_{AH}$.}
\end{figure*}

Figure 13 presents a comparison between the $\mathrm{H^+}$ and $\mathrm{O^+}$ densities, normalized to $n_0$, at time $\Omega_{cH} t= 21.8$. Both $\mathrm{H^+}$ and $\mathrm{O^+}$ ions tend to concentrate in the same regions in proximity of the separatrices. These enhanced density regions are represented in red in Figure 13; the depleted density regions are called {\em cavities} in the literature \citep{Kleva:1995,Lapenta:2010} and are visible in blue in the same plots. The formation of depleted and enhanced density areas along the separatrices is one of the signature of the guide field magnetic reconnection \citep{Lapenta:2010}. The maximum densities are almost identical for $\mathrm{O^+}$ and $\mathrm{H^+}$, while the minimum density for $\mathrm{O^+}$ is 3.6 times less than the one for $\mathrm{H^+}$. Figure 14 shows the peak densities at time $\Omega_{cH} t= 18.5$: high density areas are not symmetric with respect to the X point, and are elongated thin structures in the case of $\mathrm{H^+}$, and are thick and shorter in the case of $\mathrm{O^+}$.

\begin{figure*}
\label{densitiesGF}
\noindent
\center
\includegraphics[width=1.0\textwidth,angle=0]{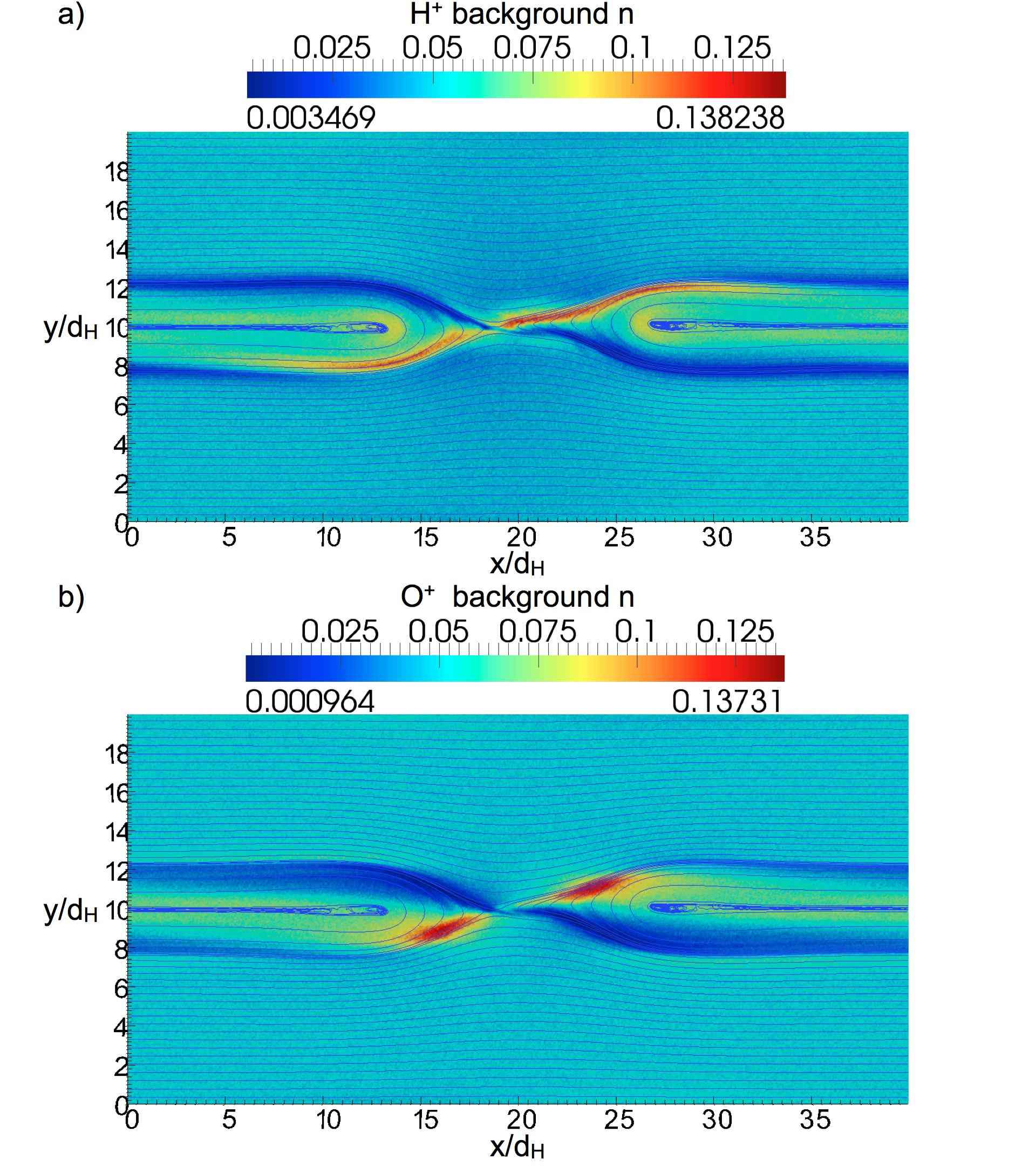} 
\caption{Densities, normalized to $n_0$, for the $\mathrm{H^+}$ and $\mathrm{O^+}$ background populations at time $\Omega_{cH} t= 21.8$ with superimposed magnetic field lines in presence of a guide field.}
\end{figure*}

\begin{figure}
\label{PeakGF}
\noindent
\center
\includegraphics[width=\columnwidth,angle=0]{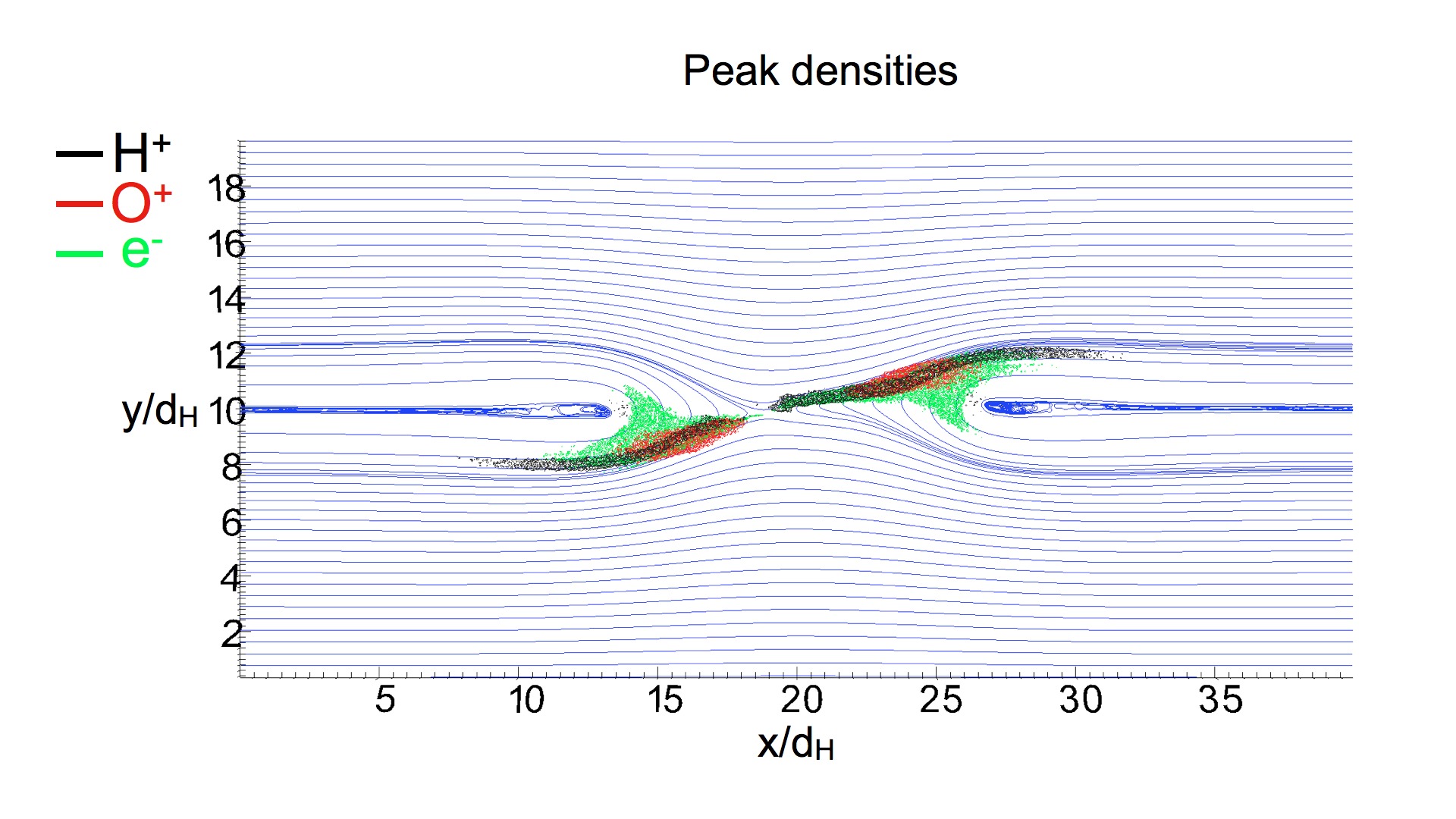}
\caption{Peak density regions for $\mathrm{H^+}$ and $\mathrm{O^+}$ and electron background species in black, red and green with superimposed magnetic field lines in presence of a guide field at time $\Omega_{cH} t= 18.5$.}
\end{figure}

Figure 15 shows a different circulation pattern for the $\mathrm{H^+}$ and $\mathrm{O^+}$ background species due the presence of a guide field.  The guide field deflects the $\mathrm{H^+}$ population in proximity of the separatrices in the inflow region and drives it to the X point. An inspection of panel b) of Figure 15 shows an $\mathrm{O^+}$ flow between separatrices: the $\mathrm{O^+}$ ions move from the cavities to enhanced density regions.

\begin{figure*}
\label{StreamingGF}
\noindent
\center
\includegraphics[width=1.0\textwidth,angle=0]{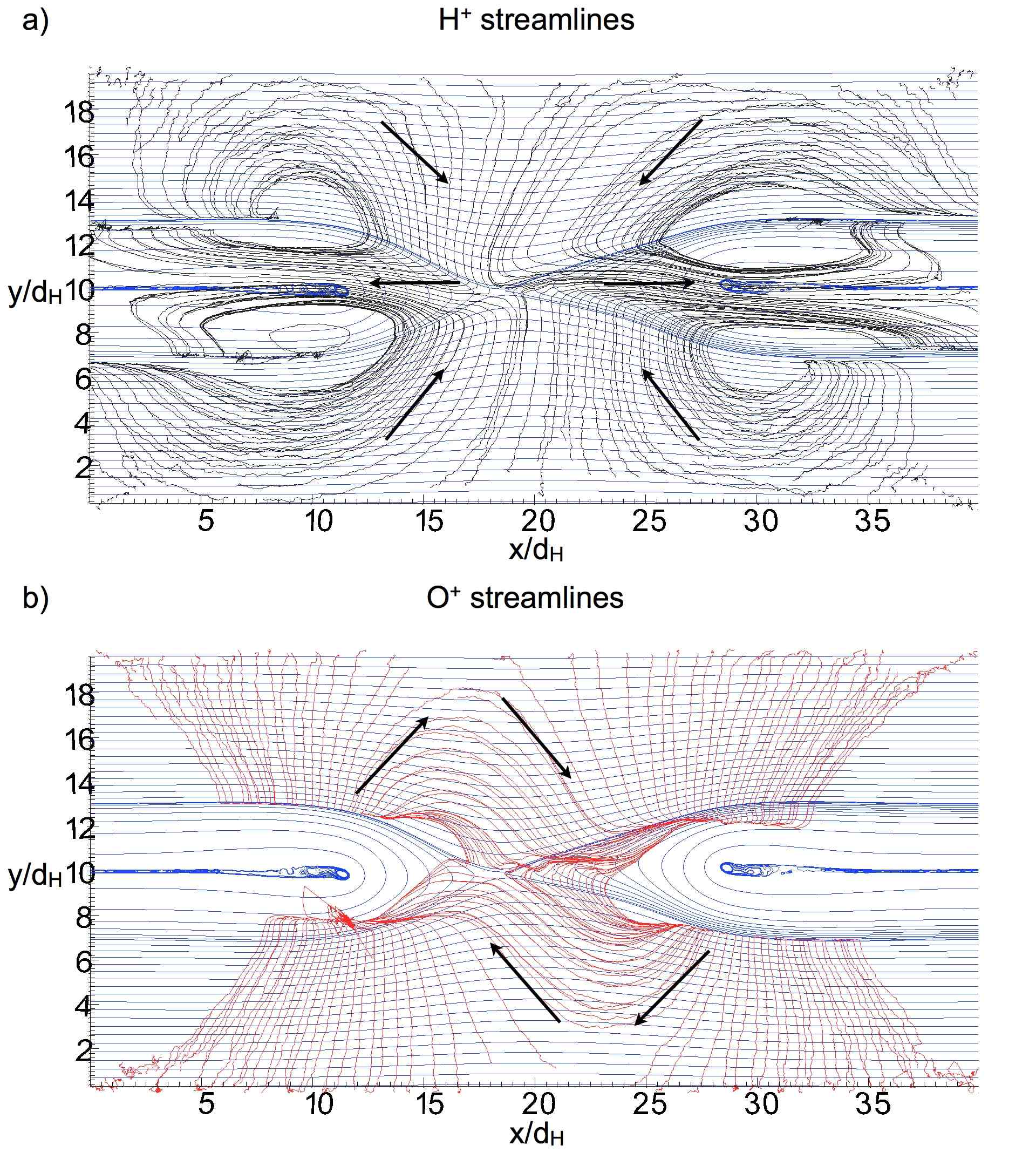} 
\caption{In-plane streamlines for the background $\mathrm{H^+}$ and $\mathrm{O^+}$ populations in black and red with superimposed magnetic lines in blue at time $\Omega_{cH} t= 21.8$ in presence of a guide field. The black arrows show the flow direction.}
\end{figure*}


\section{Discussion and conclusions}
The simulations clearly show that the presence of an $\mathrm{O^+}$ population slightly decreases the reconnection rate and magnetic reconnection speed depends mainly on the lighter $\mathrm{H^+}$ ion background species. This result is in agreement with previous Particle-in-Cell simulations \citep{Hesse:2004, Karimabadi:2010} with reduced mass ratios. However, simulations with physical mass ratio closely mimic real systems, providing realistic results that can be compared with observations. 

It has been shown that magnetic reconnection with a $\mathrm{H^+}$ current sheet occurs on the time scale of tens $\Omega_{cH}^{-1}$ \citep{Pritchett:2001} in the presented simulations. Because the $\mathrm{O^+}$ ion gyro-period is 16 times slower the $\mathrm{H^+}$ gyration, $\mathrm{O^+}$ particles rotate only once. The $\mathrm{O^+}$ population is essentially demagnetized over the time scale magnetic reconnection occurs in this simulation and its dynamics is determined by normal electric field accelerations in proximity of the reconnection fronts. This acceleration mechanism is already reported and studied in Refs. \citep{Wygant:2005,Drake:2009a}. The average magnetic force exerted on $\mathrm{O^+}$ particles is typically four times lower than the one acting on $\mathrm{H^+}$ ions, as shown in Figure 5. Differently from the $\mathrm{O^+}$ species, the background $\mathrm{H^+}$ population is effectively magnetized out of the diffusion region and decouples from the magnetic field approximately at a $d_{H}$ distance from the X point. The different dynamics of $\mathrm{H^+}$, $\mathrm{O^+}$ and electrons leads to a modified Hall current system, that produces a broader out-of-plane quadrupolar structure as reported in previous studies \citep{Shay:2004,Karimabadi:2010}. The Hall magnetic field in presence of $\mathrm{O^+}$ shows a two-scale structure: an approximately $4 d_{H}$ thick region, where the $B_z$ (core Hall magnetic field) reaches the peak values, and an extended region, where $B_z$ (tail Hall magnetic field) decreases slowly. 

Because the $\mathrm{H^+}$ and $\mathrm{O^+}$ background species undergo different acceleration mechanisms, that occur on different time scales and with different strengths, the two background species initially separate in the outflow region. The dipolarization front plasma is composed mainly of $\mathrm{H^+}$ ions, while the $\mathrm{O^+}$ species follows the $\mathrm{H^+}$ population in the outflow region. 

The reported simulations results lead to reasonable observational expectations in present and future missions. The tail Hall magnetic field is a possible signature of the background $\mathrm{O^+}$ presence during magnetic reconnection. Moreover, if $\mathrm{H^+}$ ions are current carriers and a mixed background population of $\mathrm{O^+}$ and $\mathrm{H^+}$ is present in the lobe plasma, the dipolarization fronts should be composed of a majority of $\mathrm{H^+}$ plasma, and followed by a more extended $\mathrm{O^+}$ tail. 

Finally, the effect of a guide field on the $\mathrm{O^+}$ dynamics during magnetic reconnection has been studied. These simulations are of interest for reconnection in the magnetopause with $\mathrm{O^+}$ ions \citep{Pritchett:2001,Lindstedt:2010}. It has been found that the $\mathrm{O^+}$ presence does not change the reconnected flux and therefore the pace magnetic reconnection occurs. A comparison of simulation with and without guide field reveals that a guide field reduces the reconnected flux. The typical signatures of guide field magnetic reconnection, such as outflow asymmetry and the formation of region with enhanced and depleted regions along the separatrices, have been observed in presence of a background $\mathrm{O^+}$ also. In addition, an $\mathrm{O^+}$ flow, moving from the cavities to areas with enhanced density, has been identified for the first time.

Two issues regarding to the validity of presented simulations should be finally addressed.

First, it should be discussed if the grid resolution is fine enough to capture the electron layer physics and resolve the electron scales. Figure 16 shows the local grid resolution measured in terms of electron skin depth $d_e=c/\omega_{pe}$ at time $\Omega_{cH} t= 17.5$. The electron dissipation region is properly resolved ($\Delta x/ d_e \approx 0.1$). The local grid spacing  $\Delta x /d_e$ is between 0.6 and 0.7 in the outflow area, where the species separation takes place. Thus, the electron scales are resolved by the grid in the regions of interest for this paper. However, the grid does not resolve the electron scales in the areas covered by exhaust plasma, where $\Delta x /d_e \approx 2$.
\begin{figure*}
\label{LocalGrid}
\noindent
\center
\includegraphics[width=1.0\textwidth,angle=0]{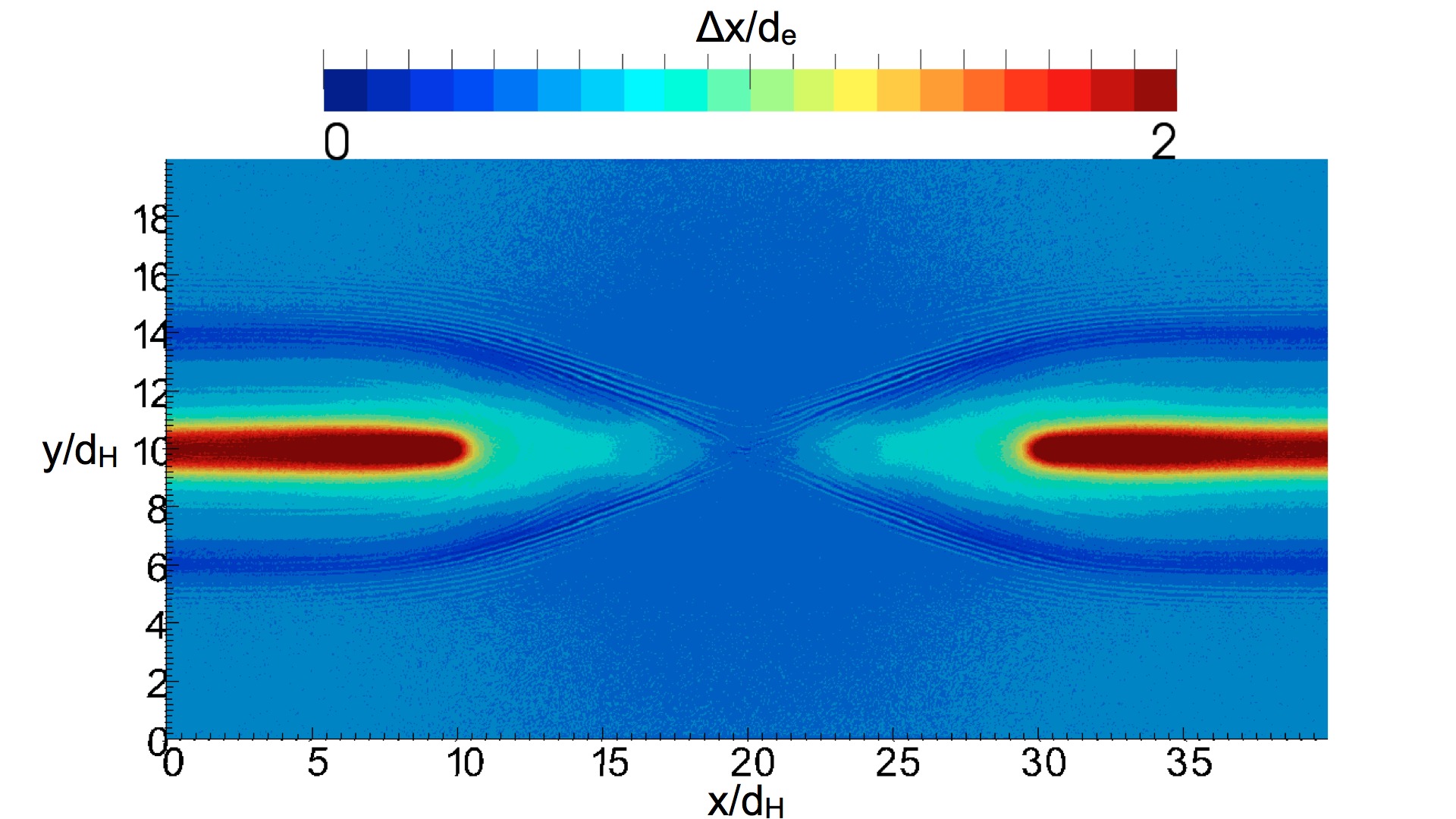} \\
\caption{Local grid spacing measured in terms of the local electron skin depth $d_e$ at time $\Omega_{cH} t= 17.5$.}
\end{figure*}

Second, the question if the $\mathrm{O^+}$ dynamics is allowed to properly couple to largest scales in a $40d_{H} \times 20d_{H} $ box, should be addressed. In fact, the size of this domain is not large enough to accommodate the whole $\mathrm{O^+}$ dissipation layer, where the $\mathrm{O^+}$ ions would eventually couple to the $\mathrm{H^+}$ species. This limitation is evident in Figure 4, where the quadrupole Hall magnetic field $B_z$ extends to the edge of the simulation box and vanishes because of the imposed boundary conditions. To answer this question, a simulation, whose domain is four times larger in each dimension, has been completed: a $160d_{H} \times 80d_{H} $ box is discretized in a $4096 \times 2048$ grid and 6.5 billion particles are used. All the other parameters, physical mass ratios included, are the same of the simulation with smaller box. From the two panels of Figure 17, representing a pseudocolor plot of the Hall magnetic $B_z$ (panel a) and the $B_z$ profile along the line $x = 68 d_H$ at time $\Omega_{cH} t= 17.46$ (panel b), it is clear that the larger simulation box includes the whole $\mathrm{O^+}$ diffusion region. The quadrupole Hall magnetic field, signature of the $\mathrm{H^+}$ and $\mathrm{O^+}$ dissipation regions, extends over a limited area and vanishes in proximity of the current layer. The analysis of other quantities, that are not reported here, shows additional evidence that the $160d_{H} \times 80d_{H} $ simulation box includes the whole $\mathrm{O^+}$ diffusion region.

\begin{figure*}
\label{DiffRegion}
\noindent
\center
\includegraphics[width=1.0\textwidth,angle=0]{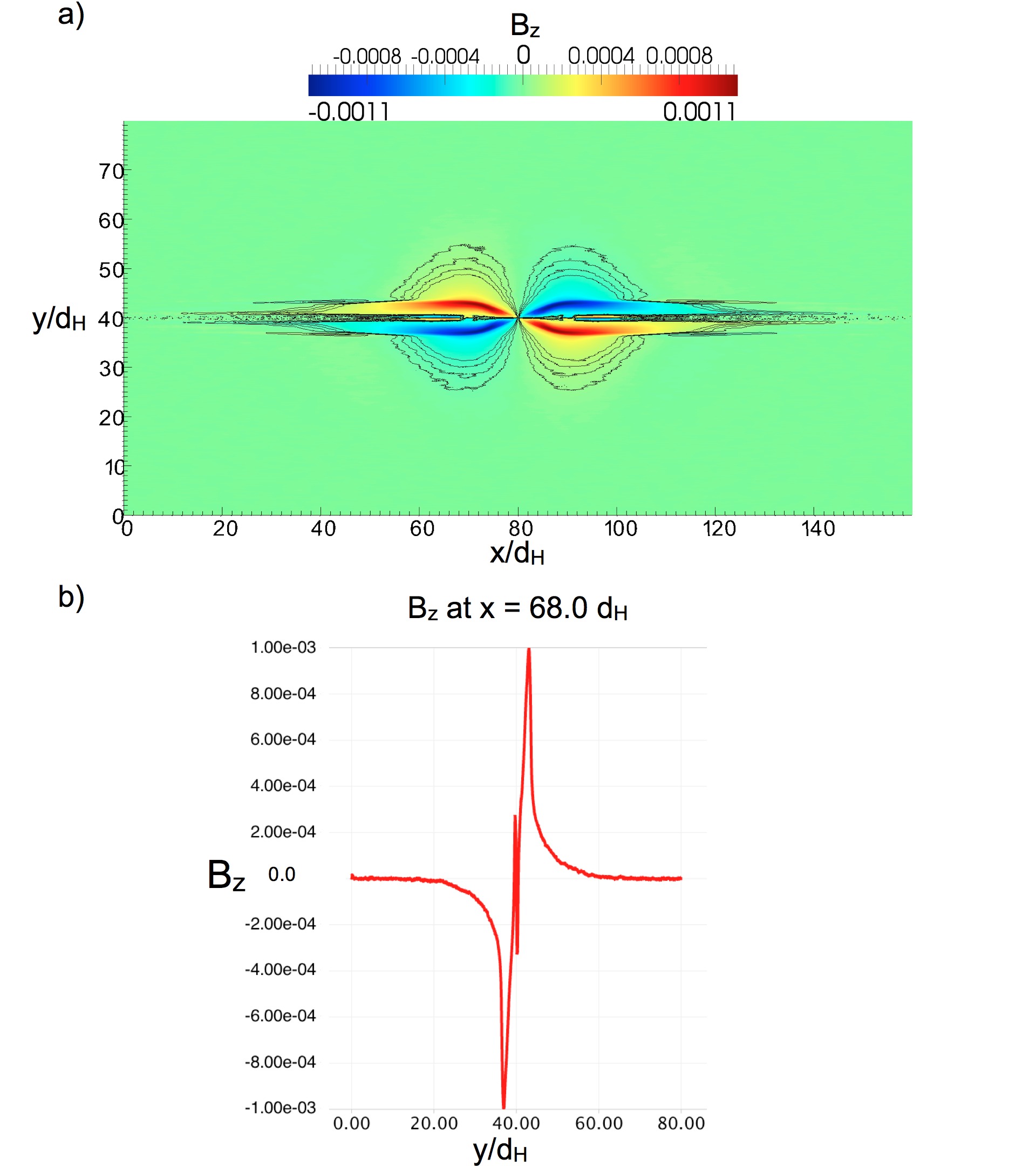} \\
\caption{Pseudocolor plot of the Hall magnetic field $B_z$ with superimposed black contour-lines to represent the non-zero areas of $B_z$ in the $160d_{H} \times 80d_{H}$ simulation box at time $\Omega_{cH} t= 17.46$ in panel a). The line profile of the Hall magnetic field $B_z$ along the line $x = 68d_H$ is shown in panel b). The two plots reveal that the Hall magnetic field vanishes in a region close to the current sheet, and the whole $\mathrm{O^+}$ diffusion region is included in the $160d_{H} \times 80d_{H}$ simulation box.}
\end{figure*}

All the features of magnetic reconnection in presence of $\mathrm{O^+}$ ions are recovered in the simulation with larger box. Magnetic reconnection starts later in time after approximately $1.5 \Omega_{cH}^{-1}$, and proceeds at the same pace of the magnetic reconnection in the smaller box. Panel a) of  Figure 18 shows a comparison of the reconnected fluxes of the two simulations. The bottom abscissa refers to the simulation with $40d_{H} \times 20d_{H} $ box, while the top abscissa to the $160d_{H} \times 80d_{H}$ simulation. A time lag of $1.5 \Omega_{cH}^{-1}$ is differentiating the two abscissae. The magnetic reconnection starts later in the larger box, the growth rate perfectly matches in the two simulations with different box size and the reconnection rate does not change with the size of the simulation domain.

Moreover, the separation of species occurs in the larger box simulation also. The panel b) of Figure 18 shows the peak densities for $\mathrm{O^+}$ and $\mathrm{H^+}$ ions in the $160d_{H} \times 80d_{H} $ simulation box, confirming that the $\mathrm{O^+}$ and $\mathrm{H^+}$ still separate in the outflow region, because of different acceleration mechanisms. In addition, the two-scale structure of the Hall magnetic field has been observed in the simulation with larger box also. Thus, after a comparison of the two simulations, it is clear that in the presented simulations there is no strong coupling of the $\mathrm{O^+}$ dynamics with the larger scales and that simulations with smaller box reproduce realistically the $\mathrm{O^+}$ behavior.

\begin{figure*}
\label{LargeBox}
\noindent
\center
\includegraphics[width=1.0\textwidth,angle=0]{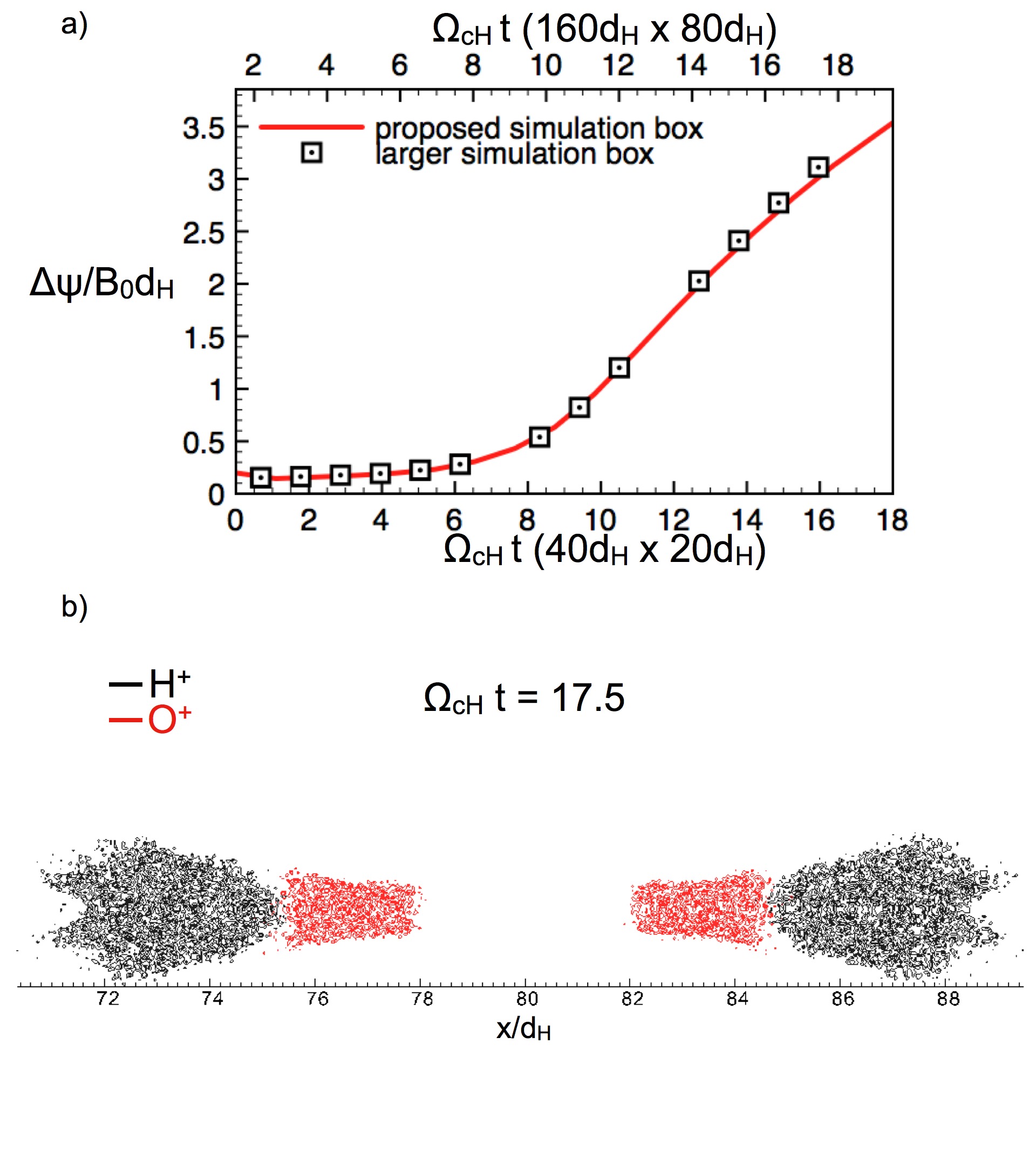} \\
\caption{Comparison of the reconnected fluxes for simulations with $40d_{H} \times 20d_{H}$ (proposed simulation box) and $160d_{H} \times 80d_{H}$ (larger simulation box)  sizes in panel a). Species separation still occurs in the $160d_{H} \times 80d_{H} $ simulation box at time $\Omega_{cH} t= 17.5$ in panel b).}
\end{figure*}


%
%
%
%
%
%

%
%
%
%

\begin{acknowledgments}
The present work is supported by the NASA MMS Grant NNX08AO84G, by the Onderzoekfonds KU Leuven (Research Fund KU Leuven) and by the European Commission's Seventh Framework Programme (FP7/2007-2013) under the grant agreement no. 263340 (SWIFF project, www.swiff.eu). Simulations were conducted on the resources of the NASA Advanced Supercomputing Division (NAS), of the NASA Center for Computational Sciences Division (NCCS) and of the Vlaams Supercomputer Centrum (VSC) at the Katholieke Universiteit Leuven.
\end{acknowledgments}

%
%
%
%
%
%
%
%
%
%

\begin{thebibliography}{29}
\expandafter\ifx\csname natexlab\endcsname\relax\def\natexlab#1{#1}\fi

\bibitem[{{\it {Birn} et~al.\/}(2004){\it {Birn}, {Thomsen}, and
  {Hesse}\/}}]{Birn:2004}
{Birn}, J., M.~{Thomsen}, and M.~{Hesse}, {Acceleration of oxygen ions in the
  dynamic magnetotail}, {\it Annales Geophysicae\/}, {\it 22\/}, 1305--1315,
  2004.

\bibitem[{{\it {Brambles} et~al.\/}(2010{\natexlab{a}}){\it {Brambles},
  {Lotko}, {Damiano}, {Zhang}, {Wiltberger}, and {Lyon}\/}}]{Brambles:2010}
{Brambles}, O.~J., W.~{Lotko}, P.~A. {Damiano}, B.~{Zhang}, M.~{Wiltberger},
  and J.~{Lyon}, {Effects of causally driven cusp $O^{+}$ outflow on the storm
  time magnetosphere-ionosphere system using a multifluid global simulation},
  {\it Journal of Geophysical Research\/}, {\it 115\/}, 0--+,
  2010{\natexlab{a}}.

\bibitem[{{\it {Brambles} et~al.\/}(2010{\natexlab{b}}){\it {Brambles},
  {Lotko}, {Zhang}, {Lyon}, and {Wiltberger}\/}}]{Brambles:2010b}
{Brambles}, O.~J., W.~{Lotko}, B.~{Zhang}, J.~{Lyon}, and M.~J. {Wiltberger},
  {Magnetospheric Sawtooth Oscillations Induced by Ionospheric Outflow}, {\it
  AGU Fall Meeting Abstracts\/}, pp. A1827+, 2010{\natexlab{b}}.

\bibitem[{{\it {Drake} et~al.\/}(2009{\natexlab{a}}){\it {Drake}, {Cassak},
  {Shay}, {Swisdak}, and {Quataert}\/}}]{Drake:2009b}
{Drake}, J.~F., P.~A. {Cassak}, M.~A. {Shay}, M.~{Swisdak}, and E.~{Quataert},
  {A Magnetic Reconnection Mechanism for Ion Acceleration and Abundance
  Enhancements in Impulsive Flares}, {\it \apjl\/}, {\it 700\/}, L16--L20,
  2009{\natexlab{a}}.

\bibitem[{{\it {Drake} et~al.\/}(2009{\natexlab{b}}){\it {Drake}, {Swisdak},
  {Phan}, {Cassak}, {Shay}, {Lepri}, {Lin}, {Quataert}, and
  {Zurbuchen}\/}}]{Drake:2009a}
{Drake}, J.~F., M.~{Swisdak}, T.~D. {Phan}, P.~A. {Cassak}, M.~A. {Shay}, S.~T.
  {Lepri}, R.~P. {Lin}, E.~{Quataert}, and T.~H. {Zurbuchen}, {Ion heating
  resulting from pickup in magnetic reconnection exhausts}, {\it Journal of
  Geophysical Research (Space Physics)\/}, {\it 114\/}, A05,111,
  2009{\natexlab{b}}.

\bibitem[{{\it {Frank} et~al.\/}(1977){\it {Frank}, {Ackerson}, and
  {Yeager}\/}}]{Frank:1977}
{Frank}, L.~A., K.~L. {Ackerson}, and D.~M. {Yeager}, {Observations of atomic
  oxygen $O^+$ in the earth's magnetotail}, {\it \jgr\/}, {\it 82\/}, 129--134,
  1977.

\bibitem[{{\it {Fujimoto} and {Nakamura}\/}(1994)}]{Fujimoto:1994}
{Fujimoto}, M., and M.~{Nakamura}, {Acceleration of heavy ions in the
  magnetotail reconnection layer}, {\it \grl\/}, {\it 21\/}, 2955--2958, 1994.

\bibitem[{{\it {Glocer} et~al.\/}(2009{\natexlab{a}}){\it {Glocer}, {T{\'o}th},
  {Gombosi}, and {Welling}\/}}]{Glocer:2009}
{Glocer}, A., G.~{T{\'o}th}, T.~{Gombosi}, and D.~{Welling}, {Modeling
  ionospheric outflows and their impact on the magnetosphere, initial results},
  {\it Journal of Geophysical Research\/}, {\it 114\/}, 5216--+,
  2009{\natexlab{a}}.

\bibitem[{{\it {Glocer} et~al.\/}(2009{\natexlab{b}}){\it {Glocer}, {T{\'o}th},
  {Ma}, {Gombosi}, {Zhang}, and {Kistler}\/}}]{Glocer:2009b}
{Glocer}, A., G.~{T{\'o}th}, Y.~{Ma}, T.~{Gombosi}, J.~{Zhang}, and L.~M.
  {Kistler}, {Multifluid Block-Adaptive-Tree Solar wind Roe-type Upwind Scheme:
  Magnetospheric composition and dynamics during geomagnetic storms - Initial
  results}, {\it Journal of Geophysical Research\/}, {\it 114\/}, 12,203--+,
  2009{\natexlab{b}}.

\bibitem[{{\it Goldman et~al.\/}(2010){\it Goldman, Lapenta, Newman, Markidis,
  and Che\/}}]{Goldman:2010}
Goldman, M., G.~Lapenta, D.~Newman, S.~Markidis, and H.~Che, {Jet deflection by
  very weak guide fields during magnetic reconnection}, {\it submitted to
  Physical Review Letters\/}, 2010.

\bibitem[{{\it {Hesse} and {Birn}\/}(2004)}]{Hesse:2004}
{Hesse}, M., and J.~{Birn}, {On the cessation of magnetic reconnection}, {\it
  Annales Geophysicae\/}, {\it 22\/}, 603--612, 2004.

\bibitem[{{\it Karimabadi et~al.\/}(2010){\it Karimabadi, Roytershteyn,
  Mouikis, Kistler, and Daughton\/}}]{Karimabadi:2010}
Karimabadi, H., V.~Roytershteyn, C.~Mouikis, L.~Kistler, and W.~Daughton,
  Flushing effect in reconnection: Effects of minority species of oxygen ions,
  {\it Planetary and Space Science\/}, {\it In Press, Corrected Proof\/}, --,
  2010.

\bibitem[{{\it {Kistler} et~al.\/}(2005)}]{Kistler:2005}
{Kistler}, L.~M., et~al., {Contribution of nonadiabatic ions to the cross-tail
  current in an $O^{+}$ dominated thin current sheet}, {\it Journal of
  Geophysical Research (Space Physics)\/}, {\it 110\/}, 6213--+, 2005.

\bibitem[{{\it {Kistler} et~al.\/}(2006)}]{Kistler:2006}
{Kistler}, L.~M., et~al., {Ion composition and pressure changes in storm time
  and nonstorm substorms in the vicinity of the near-Earth neutral line}, {\it
  Journal of Geophysical Research (Space Physics)\/}, {\it 111\/}, 11,222--+,
  2006.

\bibitem[{{\it {Kleva} et~al.\/}(1995){\it {Kleva}, {Drake}, and
  {Waelbroeck}\/}}]{Kleva:1995}
{Kleva}, R.~G., J.~F. {Drake}, and F.~L. {Waelbroeck}, {Fast reconnection in
  high temperature plasmas}, {\it Physics of Plasmas\/}, {\it 2\/}, 23--34,
  1995.

\bibitem[{{\it Lapenta et~al.\/}(2010){\it Lapenta, Markidis, Divin, Goldman,
  and Newman\/}}]{Lapenta:2010}
Lapenta, G., S.~Markidis, A.~Divin, M.~Goldman, and D.~Newman, Scales of guide
  field reconnection at the hydrogen mass ratio, {\it Physics of Plasmas\/},
  {\it 17\/}, 082,106, 2010.

\bibitem[{{\it {Lindstedt} et~al.\/}(2010){\it {Lindstedt}, {Khotyaintsev},
  {Vaivads}, {Andr{\'e}}, {Nilsson}, and {Waara}\/}}]{Lindstedt:2010}
{Lindstedt}, T., Y.~V. {Khotyaintsev}, A.~{Vaivads}, M.~{Andr{\'e}},
  H.~{Nilsson}, and M.~{Waara}, {Oxygen energization by localized perpendicular
  electric fields at the cusp boundary}, {\it \grl\/}, {\it 37\/}, 9103--+,
  2010.

\bibitem[{{\it Markidis et~al.\/}(2010){\it Markidis, Lapenta, and
  Rizwan-uddin\/}}]{Markidis:2010}
Markidis, S., G.~Lapenta, and Rizwan-uddin, Multi-scale simulations of plasma
  with i{PIC}3D, {\it Mathematics and Computers in Simulation\/}, {\it 80\/},
  1509 -- 1519, 2010.

\bibitem[{{\it Mouikis et~al.\/}(2010){\it Mouikis, Kistler, Liu, Klecker,
  Korth, and Dandouras\/}}]{Mouikis:2010}
Mouikis, C.~G., L.~M. Kistler, Y.~H. Liu, B.~Klecker, A.~Korth, and
  I.~Dandouras, $H^{+}$ and $O^{+}$  content of the plasma sheet at 15-19 Re as
  a function of geomagnetic and solar activity, {\it J. Geophys. Res.\/}, {\it
  115\/}, 2010.

\bibitem[{{\it {Pritchett}\/}(2001)}]{Pritchett:2001}
{Pritchett}, P.~L., {Geospace Environment Modeling magnetic reconnection
  challenge: Simulations with a full particle electromagnetic code}, {\it
  \jgr\/}, {\it 106\/}, 3783--3798, 2001.

\bibitem[{{\it {Ricci} et~al.\/}(2002){\it {Ricci}, {Lapenta}, and
  {Brackbill}\/}}]{Ricci:2002}
{Ricci}, P., G.~{Lapenta}, and J.~U. {Brackbill}, {GEM reconnection challenge:
  Implicit kinetic simulations with the physical mass ratio}, {\it \grl\/},
  {\it 29\/}, 230,000--1, 2002.

\bibitem[{{\it {Ricci} et~al.\/}(2004){\it {Ricci}, {Brackbill}, {Daughton},
  and {Lapenta}\/}}]{Ricci:2004}
{Ricci}, P., J.~U. {Brackbill}, W.~{Daughton}, and G.~{Lapenta}, {Collisionless
  magnetic reconnection in the presence of a guide field}, {\it Physics of
  Plasmas\/}, {\it 11\/}, 4102--4114, 2004.

\bibitem[{{\it Shay and Swisdak\/}(2004)}]{Shay:2004}
Shay, M.~A., and M.~Swisdak, Three-species collisionless reconnection: Effect
  of $O^{+}$ on magnetotail reconnection, {\it Phys. Rev. Lett.\/}, {\it 93\/},
  175,001, 2004.

\bibitem[{{\it {Swisdak} et~al.\/}(2005){\it {Swisdak}, {Drake}, {Shay}, and
  {McIlhargey}\/}}]{Swisdak:2005}
{Swisdak}, M., J.~F. {Drake}, M.~A. {Shay}, and J.~G. {McIlhargey}, {Transition
  from antiparallel to component magnetic reconnection}, {\it Journal of
  Geophysical Research (Space Physics)\/}, {\it 110\/}, 5210--+, 2005.

\bibitem[{{\it {Wilken} et~al.\/}(1995){\it {Wilken}, {Zong}, {Daglis}, {Doke},
  {Livi}, {Maezawa}, {Pu}, {Ullaland}, and {Yamamoto}\/}}]{Wilken:1995}
{Wilken}, B., Q.~G. {Zong}, I.~A. {Daglis}, T.~{Doke}, S.~{Livi}, K.~{Maezawa},
  Z.~Y. {Pu}, S.~{Ullaland}, and T.~{Yamamoto}, {Tailward flowing energetic
  oxygen ion bursts associated with multiple flux ropes in the distant
  magnetotail: GEOTAIL observations}, {\it \grl\/}, {\it 22\/}, 3267--3270,
  1995.

\bibitem[{{\it {Wiltberger} et~al.\/}(2010){\it {Wiltberger}, {Lotko}, {Lyon},
  {Damiano}, and {Merkin}\/}}]{Wiltberger:2010}
{Wiltberger}, M., W.~{Lotko}, J.~G. {Lyon}, P.~{Damiano}, and V.~{Merkin},
  {Influence of cusp $O^{+}$ outflow on magnetotail dynamics in a multifluid
  MHD model of the magnetosphere}, {\it Journal of Geophysical Research\/},
  {\it 115\/}, 0--+, 2010.

\bibitem[{{\it {Winglee} et~al.\/}(2002){\it {Winglee}, {Chua}, {Brittnacher},
  {Parks}, and {Lu}\/}}]{Winglee:2002}
{Winglee}, R.~M., D.~{Chua}, M.~{Brittnacher}, G.~K. {Parks}, and G.~{Lu},
  {Global impact of ionospheric outflows on the dynamics of the magnetosphere
  and cross-polar cap potential}, {\it Journal of Geophysical Research\/}, {\it
  107\/}, 1237--+, 2002.

\bibitem[{{\it {Wygant} et~al.\/}(2005)}]{Wygant:2005}
{Wygant}, J.~R., et~al., {Cluster observations of an intense normal component
  of the electric field at a thin reconnecting current sheet in the tail and
  its role in the shock-like acceleration of the ion fluid into the separatrix
  region}, {\it Journal of Geophysical Research (Space Physics)\/}, {\it
  110\/}, 9206--+, 2005.

\bibitem[{{\it {Zong} et~al.\/}(1998)}]{Zong:1998}
{Zong}, Q., et~al., {Energetic oxygen ion bursts in the distant magnetotail as
  a product of intense substorms: Three case studies}, {\it \jgr\/}, {\it
  103\/}, 20,339--20,364, 1998.

\end{thebibliography}








%
%

\end{article}




%
%
%
%
%
%


\end{document}